\documentclass{aa}  

\usepackage{graphicx,url}
\usepackage{txfonts,xcolor}

\def\jsix{SDSS~J0100+2802}
\def\jfive{SDSS~J0306+1853}
\def\b2{B2~1023+25}

\begin{document}

   \title{Jetted radio-quiet quasars at z>5}

   \subtitle{}

   \author{T.\ Sbarrato\inst{1}, G.\ Ghisellini\inst{1}, G.\ Giovannini\inst{2,3},
          \and
          M.\ Giroletti\inst{2}
          }

   \institute{INAF--Osservatorio Astronomico di Brera, via E.\ Bianchi 46, 23807 Merate (LC), Italy\\
              \email{tullia.sbarrato@inaf.it}
         \and
   INAF--Istituto di Radioastronomia, via P.\ Gobetti 101, 40129 Bologna, Italy \and
   University of Bologna, Department of Physics and Astronomy, via P.\ Gobetti 93, 40129 Bologna, Italy
             }

   \date{}

 
  \abstract{
  We report on the JVLA observations of three high-redshift active galactic nuclei (AGNs) that have black hole masses estimated to be among the largest known.
  Two of them, \jsix\ and \jfive\ at redshift 6.326 and 5.363, respectively, 
  are radio-quiet AGNs according to the classic definition, while 
  the third (\b2\ at $z=5.284$) is a powerful blazar. 
  The JVLA data clearly show a radio structure in the first source
  and a radio emission with a relatively steep radio spectrum in the second one,
  indicating the presence of a radio jet and a diffuse component.
  Therefore, being radio-quiet does not exclude the presence of a powerful 
  relativistic jet, which has important consequences on population studies and on the ratio between jetted and non-jetted AGNs.
  We can estimate the viewing angle of these jets, and this allows us to find,
  albeit with some uncertainty, the density of black holes with a mass 
  in excess of $10^{10} M_\odot$ at high redshifts. 
  We found that their density in jetted AGNs is very large in the redshift bin 5--6 and comparable with the overall AGN population of the same optical luminosity.
  Jets might thus play a crucial role in the fast formation and evolution 
  of the most massive black holes in the early Universe. 
  They are more common than what is expected from wide radio surveys with milliJansky flux sensitivity. 
  Deeper JVLA or very-long-baseline interferometry observations are key to discovering a possible relativistic 
  jet population hiding in plain sight at very high redshift.
  The discovery of powerful relativistic jets associated with the most massive
  black holes in the early Universe revives the question: is the jet instrumental for a rapid growth of the black hole or,
  instead, is the black hole mass the main driver for the jet formation?
}

   \maketitle
%

\section{Introduction}
\label{sec:intro}

The number of known quasars at redshifts larger than 5 has been dramatically increasing in the last few years, following their first discovery more than 20 years ago \citep{fan99}. 
Currently, $\sim300$ sources have been discovered. 
The few for which a virial-based black hole mass has been derived host supermassive black holes (SMBHs) with $M_{\rm BH}\geq10^9M_\odot$. 
At such high redshift, host galaxies show intense star forming activity, with star formation rates of $100-2500M_\odot/{\rm yr}$ \citep{decarli18,shao19}, suggesting that in that phase black holes outgrow their host galaxies \citep{venemans16, neeleman21}.

The formation and life cycle of early active SMBHs have been long-standing puzzles since their first observations \citep[see the reviews by][and references therein]{volonteri12,inayoshi20}. 
The expected Eddington-limited accretion through an $\alpha$-disc \citep{shakura73,novikov73} would imply a seed black hole of more than $10^4M_\odot$. 
To form such a massive seed, direct collapse from a massive gas cloud would be needed, which would likely cool and fragment before collapsing into a compact object. 
Assembling $\geq10^9M_\odot$ black holes from more realistically expected seeds of $\sim100M_\odot$ instead would imply breaking the Eddington limit.
Both solutions are strongly affected by issues and difficulties. 
Finding and studying more and more massive quasars at $z>5$ will keep challenging the current formation models, but it will also allow us to dig deeper into this subject.

Among these massive sources, some are particularly interesting: SDSS J010013.02+280225.8 (hereafter \jsix) and SDSS J030642.51+185315.8 (\jfive) are the only two known quasars with masses larger than $10^{10}M_\odot$ at redshifts higher than 5.
Specifically, \jsix\ is at redshift $z = 6.326$, and its mass is $M \simeq 1.2 \times 10^{10}M_\odot$ \citep{wu15}, while J0306+1853 is at $z = 5.363$ with a mass of $M \simeq 1.1 \times 10^{10}M_\odot$ \citep{wu15,wang15}. 
They are definitely the most challenging objects: how is it possible to build such massive black holes in less than $\sim 1.1$ Gyr (i.e.\ the age of the Universe at $z = 5$)?

\jsix\  was selected by \citet{wu15} as a high-redshift quasar candidate from a photometric cross-matched sample of the Sloan Digital Sky Survey \citep[SDSS;][]{york00}, the Two Micron All-Sky Survey \citep[2MASS;][]{skrutskie06}, and the {\it Wide-Field Infrared Survey Explorer} \citep[WISE;][]{wright10} because of its red optical colours and bright IR detection. 
Spectroscopic confirmation was obtained with the Lijiang 2.4 m telescope, the Multiple Mirror Telescope (MMT), and the Large Binocular Telescope (LBT). 
The optical--near-IR spectrum obtained with LBT, Magellan, and Gemini shows CIV and MgII lines. 
The latter allowed Wu and collaborators to obtain a virial-based black hole mass estimate for this source. 
\citet{fujimoto20} suggest that \jsix\ is likely gravitationally lensed with a magnification factor of $\mu\sim450$. 
On the basis of the Ly$\alpha$-transparent proximity zone, \citet{davies20} rule out magnifications of $\mu>4.9$ at 95\% and of $\mu>100$ conclusively. 
Finally, \citet{pacucci20} performed a consistency check on the magnification results with the quasar luminosity function, finding that if J0100+2802 is magnified, then all $z>6$ quasars in SDSS should be as well. 

\jfive\  was selected by \citet{wang15} as a high-redshift quasar candidate based on SDSS and {\it WISE} photometric data because of its red colour and IR brightness. 
Spectroscopic confirmation was obtained with the Lijiang 2.4 m telescope.
The optical--near-IR spectrum obtained with the Magellan telescope shows CIV and MgII lines. 
Wang and collaborators derived virial-based black hole mass estimates from both emission lines.

Relativistic jets have been suggested to play a role in such fast formation and accretion \citep{jolley08,ghisellini13,regan19}. 
Feedback from the jet is also largely regarded as crucial in regulating the common evolution of black holes and their host galaxies. 
The presence of a relativistic jet is generally associated with strong radio emission, and the radio-loud fraction in a quasar population is commonly used to trace the occurrence of jetted sources. 
The radio-loud fraction in $z\sim6$ quasars seems consistent with or lower than the same value in the local Universe \citep{liu21}. 
However, when aligned sources are considered to trace the jetted population (i.e.\ when blazars classified through high-energy properties are used as proxies for all jetted sources), the jetted fraction significantly increases at $z>3.5$ \citep{ghisellini10,sbarrato15,sbarrato21}. 
Therefore, either we are losing misaligned sources in radio and/or optical catalogues or we are misclassifying a fraction of blazars \citep{cao17}. 
It is already well known that the cosmic microwave background (CMB) efficiently quenches extended radio emission, resulting in a deficit of observed misaligned jetted sources \citep{bassett04,ghisellini14a,ghisellini15,napier20}.
An over-obscuration of quasar nuclei could also play a role in our inability to find and classify the misaligned population \citep{ghisellini16,fan20}. 
Jets, in fact, might be able to pierce the surrounding dust, making aligned jetted quasars more likely to be visible in optical light\footnote{ 
In this case, non-jetted quasars would be underestimated as well, complicating the picture.}.
For this reason, the radio-loud fraction alone should not be used as a reliable proxy for the number of jetted sources. 
A thorough study of broadband properties is needed to derive the presence of a jet at high redshift.

Although the jet might help the SMBH to accrete more and faster, in the literature there is no observational signature of jet presence in \jsix\ or \jfive. Neither of these two objects is in fact included in any large radio catalogue, such as the Faint Images of the Radio Sky at Twenty Centimeters \citep[FIRST;][]{becker94} or the NRAO VLA Sky Survey \citep[NVSS;][]{condon98}. 
\cite{wang16,wang17} observed \jsix\ radio continuum and line emissions with the Very Large Array (VLA) and the Very Long Baseline Array (VLBA), concluding that it cannot be due only to star formation activity; however, they did not infer the presence of a jet.
Since the extended emission of these two sources is likely quenched by the CMB, though, we chose to rely on observations with a microJansky flux limit and search with a deeper sensitivity. In the following sections, we present the observations we performed on these two sources with the JVLA.

Along with these two extreme sources (in terms of their mass and luminosity), we also performed observations on a known high-$z$, $M>10^9M_\odot$ jetted source with its jet aligned to our line of sight: SDSS J102623.61+254259.5 (hereafter B2~1023+25, $z=5.28$). 
We aimed at comparing known extremely massive jetted sources in the same redshift bin. 
Unfortunately, no $\sim10^{10}M_\odot$ blazar has yet been detected at $z>5$; therefore, we chose a very well-studied source with the largest mass possible.
This source was selected as a potential blazar candidate from the SDSS spectroscopic sample cross-correlated with FIRST because of its strong radio flux ($F_{\rm r}> 100$mJy) and extreme radio-loudness \citep{sbarrato13a}. \cite{sbarrato12} observed it in the X-rays and managed to classify it as a blazar thanks to its strong and hard X-ray flux. 
This classification was further confirmed thanks to the Nuclear Spectroscopic Telescope Array \citep[{\it NuSTAR};][]{sbarrato13b} and very-long-baseline interferometry (VLBI) observations.
The latter also allowed \cite{frey15} to observe the first jet proper motion at $z>5$. B2~1023+25 has no single-epoch virial-based mass estimates, but its mass falls in the $2-5\times10^9M_\odot$ range, according to spectral energy distribution (SED) fitting. 

In this paper we present our study on the jet and accretion features of the three sources, with the aim of studying the evolution of the most massive sources in the early Universe. Section \ref{sec:radio} presents the JVLA radio observations of the three sources. Section \ref{sec:accr} shows the comparison of the available IR-optical-UV data with different accretion disc emission models in order to assess the masses and accretion regimes of the sources. Section \ref{sec:jet} discusses the broadband SED modelling, aimed at understanding the presence and features of relativistic jets. Section \ref{sec:disc} discusses the implication of our findings in the picture of quasar distribution and evolution in the early Universe, and Sect. \ref{sec:concl} presents our conclusions.

In this work we adopt a flat cosmology with 
$H_0 = 70$ km s$^{-1}$ Mpc$^{-1}$ and $\Omega_{\rm M}=0.3$. The radio spectral index $\alpha$ is defined such that the flux density $S_\nu$ is proportional to $\nu^{-\alpha}$.

\section{Radio observations}
\label{sec:radio}

   \begin{figure*}
   \centering
   \includegraphics[height=0.5\hsize]{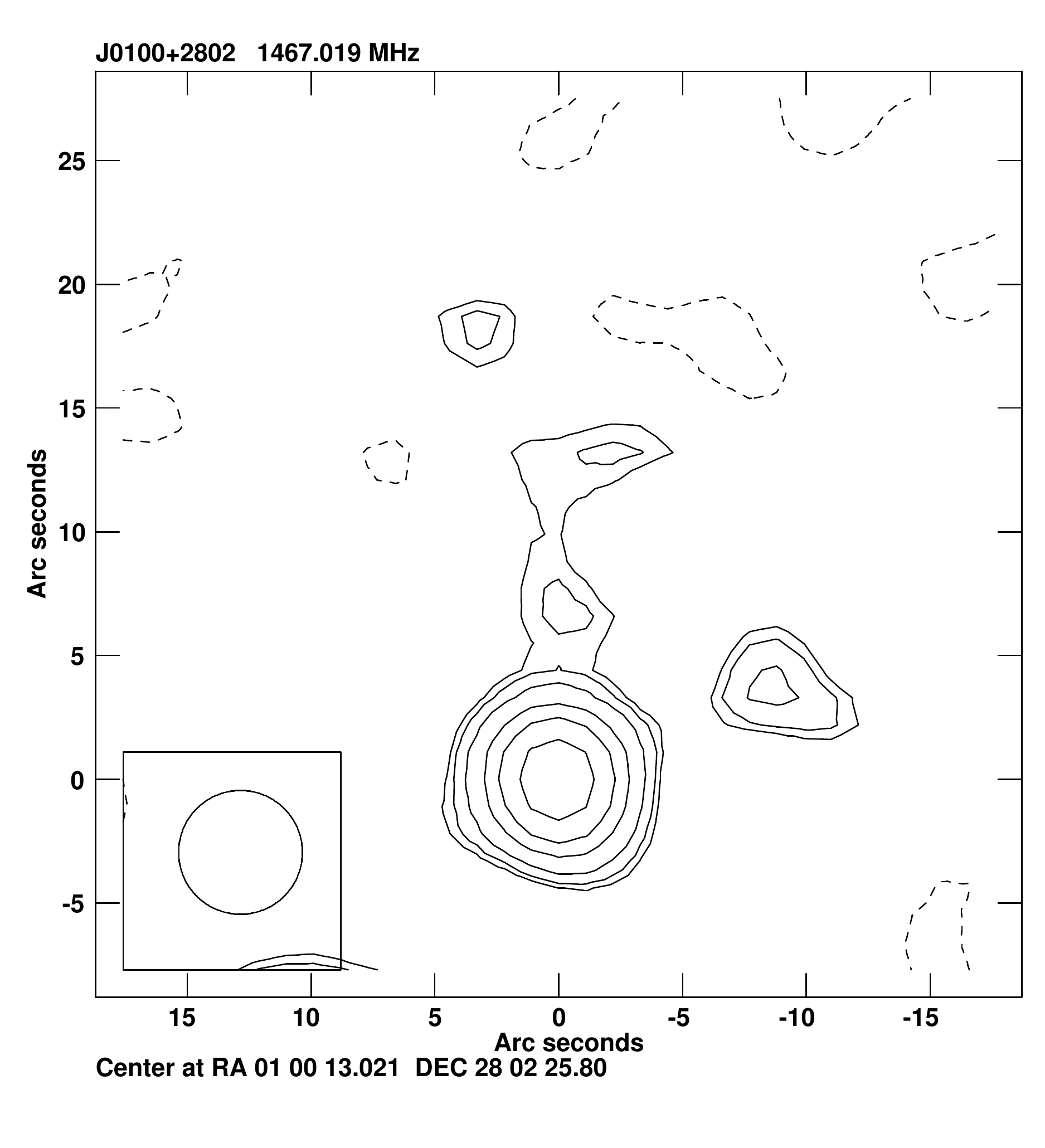}
   \includegraphics[height=0.5\hsize]{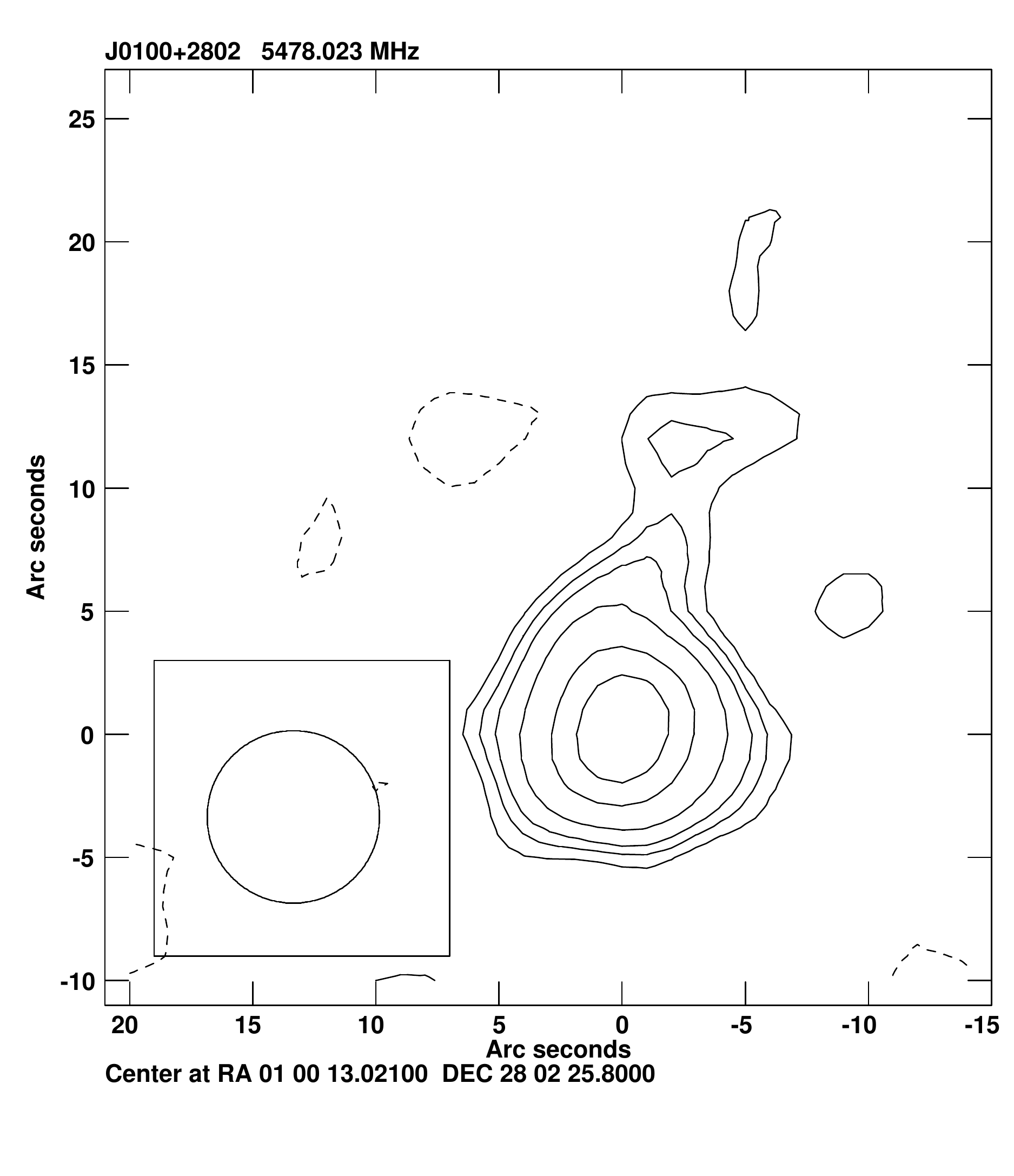}
   \caption{Radio images of \jsix{} obtained with JVLA. Left panel: 1.5 GHz image. The HPBW is 5 arcsec; surface brightness levels are -20, 15 20 30 50 70 100 $\mu$Jy/beam. Right panel: 5.5 GHz.\ The HPBW is 7 arcsec; levels are -10, 10 15 20 30 50 70 100 200 300 500 $\mu$Jy/beam
                }
    \label{fig:0100-vla}%
    \end{figure*}

\subsection{Data reduction}

\begin{table*}
\caption{Observed results. \newline\textbf{Notes}. Column 1: source name; Col. 2: measured redshift; Col. 3: angular to linear scale conversion factor; 
         Col. 4: luminosity distance; Cols. 5 and 6: observing frequency and angular resolution; 
         Col. 7: measured noise level in images; Col. 8: nuclear component flux density, the peak of a Gaussian fit; Col. 9: nuclear radio power (logarithm), no K-correction applied; Col. 10: total flux density; Col. 11: total radio power (logarithm), no K-correction applied; 
         Col. 12: source size in our images.
         }
\label{tab:radio-obs} 
\centering\small
\begin{tabular}{llcccccccccr}
\hline \hline
\noalign{\smallskip}
Name    &  z  & kpc/"  &  D$_l$ & Freq & HPBW  &  noise    & S$_c$ & Log P$_c$ & S$_t$ & LogP$_t$ & size \\
        &     &        &  Mpc   & GHz  & "     &  mJy/beam & mJy   & [W/Hz]    & mJy   & [W/Hz] & " \\ 
\noalign{\smallskip}
\hline
\noalign{\smallskip}
J0100+2802 & 6.326 & 5.556 & 61073 & 1.5 & 5 & 0.012 & 0.062$\pm$0.012 & 25.45 & 0.154$\pm$0.012 & 25.84  &13\\
           &       &       &       & 5.0 & 7 & 0.005 & 0.091$\pm$0.005 & 25.61 & 0.113$\pm$0.005 & 25.70  &13\\
J0306+1853 & 5.363 & 6.062 & 50628 & 1.5 & 4 & 0.013 & 0.252$\pm$0.014 & 25.89 & 0.252$\pm$0.014 & 25.89   &<3\\ 
           &       &       &       & 5.0 & 4 & 0.0045 & 0.088$\pm$0.005 & 25.43 & 0.088$\pm$0.005 & 25.43  &<3\\
B2 1023+25 & 5.284 & 6.109 & 49756 & 1.5 & 0.9 & 0.09  & 209.5$\pm$4.1 & 28.79 & 209.5$\pm$4.1 & 28.79  &<0.3\\
           &       &       &       & 5.5 & 1.0 & 0.013 & 112.3$\pm$2.2 & 28.52 & 112.3$\pm$2.2 & 28.52 &<0.3\\
\noalign{\smallskip}
\hline
\noalign{\smallskip}
\end{tabular}
\end{table*}

\subsubsection{\jsix{}}

We observed this source with the JVLA in the L band ($\lambda=$20cm) and the C band ($\lambda=$6cm).
Observations in the L band were in B configuration, with
a total bandwidth of 1 GHz (from 1 to 2 GHz). The source was observed in
seven scheduling blocks (SBs), each 45 minutes long, in the period May--August 2016.
Observations in the C band were in C configuration, with a total bandwidth of 2 GHz
(from 4.5 to 6.5 GHz). The source was observed in two 40-minute-long SBs on
January 31, 2016. At both frequencies the flux density and bandpass
calibrator was 3C 48, and J0119+3210 was the phase calibrator. 3C 48 was
observed once in each SB, and the phase calibrator was observed two or three times.

Data reduction was done using the AIPS package, applying the standard technique
suggested by the AIPS cookbook. A large fraction of data (about 15\%) in the L band were flagged because of strong interferences; in particular,
the first SB was not used because of the strong interferences that also
affected calibrator sources. In the C band, only about 5\% of data
suffered this problem. Each SB was calibrated separately. No
self-calibration was possible in the C band because of the
too-low flux density of sources in the field; in the L band we did two cycles of
phase-only self-calibration, 
exploiting a few moderately bright radio sources present in the large L-band primary beam.
Images from calibrated data were obtained using the AIPS package for each SB.

At the end, we combined all the $(u,v)$ data at each frequency to obtain the final images.
Because of the good quality of the data and the low noise level in the L-band
images, it was necessary
to map a large field and a few additional outer, smaller fields, using the AIPS task
IMAGR to clean many distant sources whose sidelobes were present in the central image region.
In C-band observations, a single field of 1024 x 1024 pixels was enough to obtain
a high quality image.
In Table \ref{tab:radio-obs} we report the image and source parameters.
The noise level in all images was estimated using the AIPS task IMEAN in a large area near the map centre with no visible radio sources. The core flux density is the peak of a Gaussian fit of the inner region (AIPS task JMFIT). The flux density of the extended jet feature was computed within the region visible in Fig. 1 using the AIPS task TVSTAT; the total flux is the sum of the extended and nuclear emission.
\subsubsection{\jfive{}}

Observations were carried out as for the previous source and during the same time range in the L band, as well as on February 2 and 3, 2016, in the C band. The gain and bandpass calibrator was 3C 48, and J0319+1901 was the phase calibrator.
The data calibration and reduction was similar, and the problems
were the same as for the previous source. The source is unresolved in our images. Results in Table \ref{tab:radio-obs} have been obtained with a Gaussian fit.

\subsubsection{\b2{}}

We observed this source in the L and C bands. In the L band we observed in A configuration, with
a total bandwidth of 1 GHz (from 1 to 2 GHz), for 30
minutes on April 12, 2014.
The observation in the C band was in B configuration, with a total bandwidth of
2 GHz (from 4.5 to 6.5 GHz). The source was observed for 30 minutes on
October 24, 2013. At both frequencies the gain and bandpass
calibrator was 3C 286, and J1013+2649 was the phase calibrator.
The data were self-calibrated in phase and amplitude. The source is unresolved in our images. Results in Table \ref{tab:radio-obs} have been obtained with a Gaussian fit.

\subsection{Results in radio band}

J0100+2802 shows, in images in the L and C band, a dominant nuclear emission with a
faint jet-like emission in the northern direction. In the L-band image the
structure has a surface brightness at the $1.5 \sigma$ level; however, the same
structure is visible in the C-band image (see Fig. \ref{fig:0100-vla}) at a more than
$3 \sigma$ level. Therefore, we conclude that it is a real one-sided jet emission.
Its length is $\sim12-14$ arcsec, which corresponds to $\sim70$ kpc. 
The core is self-absorbed
($\alpha^{1.5}_{5.0} = -0.31 \pm 0.1$), and the jet-like structure shows a steep spectrum
with $\alpha = 1.18 \pm 0.15$ (Table \ref{tab:radio-obs}).

At about 10" west of the core, we note a faint, slightly resolved emission present in images at 1.5 and 5 GHz. Its spectrum is steep; we investigated a possible optical identification of this source, but no optical emission is present in this region. It could be diffuse emission related to our target source or an unrelated emission on the line of sight. Deeper radio and optical data are necessary to investigate this emission. 

\jsix{} was observed with the JVLA at 3\,GHz \citep{wang16} and with the VLBA at 1.5\,GHz \citep{wang17}.
The reported JVLA total flux density is $104.5\pm3.1\,{\rm \mu Jy}$, 
lower than our flux density at 5 GHz, suggesting a possible source variability. 
The jet feature is not present in \citet{wang16} deep image because its surface brightness is too low to be detected in their high angular resolution -- half power beam width (HPBW) = 0.65 $\times$ 0.56 arcsec -- images. 
In the VLBA image by \cite{wang17}, 
a single, slightly extended component is present with a flux density of  
$64.6\pm9.0\,{\rm\mu Jy}$ at high angular resolution and with $91\pm17\,{\rm\mu Jy}$ in a low-resolution map.
The source flux density in the VLBA data is in the same range as our flux density estimates of the core 
($62{\rm\mu Jy}$ at 1.5 and $91{\rm\mu Jy}$ at 5 GHz), 
and the inverted spectrum is in agreement with the small VLBA size of the nuclear component.

J0306+1853 is unresolved in the L and C bands.  It corresponds to a
radio size smaller than 20 kpc; however, the spectral index is 0.87 $\pm$ 0.06,  suggesting
the presence of a sub-arcsecond structure. The lack of an extended jet structure
could be a result of the too low brightness due to the high inverse Compton losses.

B2 1025+23 is unresolved in images in the L and C band.  It corresponds to a
radio size smaller than 2 kpc with a spectral index of 0.48 $\pm$ 0.02. This source was
observed in VLBI mode with the European VLBI Network (EVN) by \cite{frey15}. It shows a core-dominated one-sided jet with a few substructures. By comparison with previous
observations, \cite{frey15} measured a super-luminal proper motion and,
using multi-frequency observations, constrained the jet orientation to
$3^\circ$ with respect to the line of sight, with a bulk 
Lorentz factor of $\Gamma=13$, which is consistent with results based on X-ray data and broadband SED fitting \citep{sbarrato12,sbarrato13b}.

\subsection{Discussion}

The high radio power of the source \jsix\  and its radio spectral properties are strong evidence of active galactic nucleus (AGN) activity with a SMBH producing relativistic jets in the source inner region. Moreover, in this source we detected, for the first time, a diffuse emission in the northern direction, whose extension requires the presence of a relativistic jet to transport the radio plasma from the core to this structure. The detection of a nuclear emission with VLBA data \citep{wang17} confirms this scenario and is not in agreement with a radio emission related to star formation activity or strong winds \citep[see][]{panessa19}.

The source \jfive\ was not observed with a VLBI array, and in JVLA images it shows a point-like image. However, its radio power and the relatively steep spectral index (0.87) suggests that a sub-arcsecond AGN jet structure is also present in this source.
The source B2~1023+25 shows a clear relativistic jet from VLBI images. 

We can conclude that all sources have an  AGN jet morphology, and next we estimate the possible jet orientation and velocity to derive the relativistic Doppler factor and the intrinsic source parameters. Their radio power and properties are the same as those of sources in the unified model scenario.

\begin{figure}
   \vskip -0.4 cm
   \hskip -0.1 cm
   \includegraphics[width=10cm]{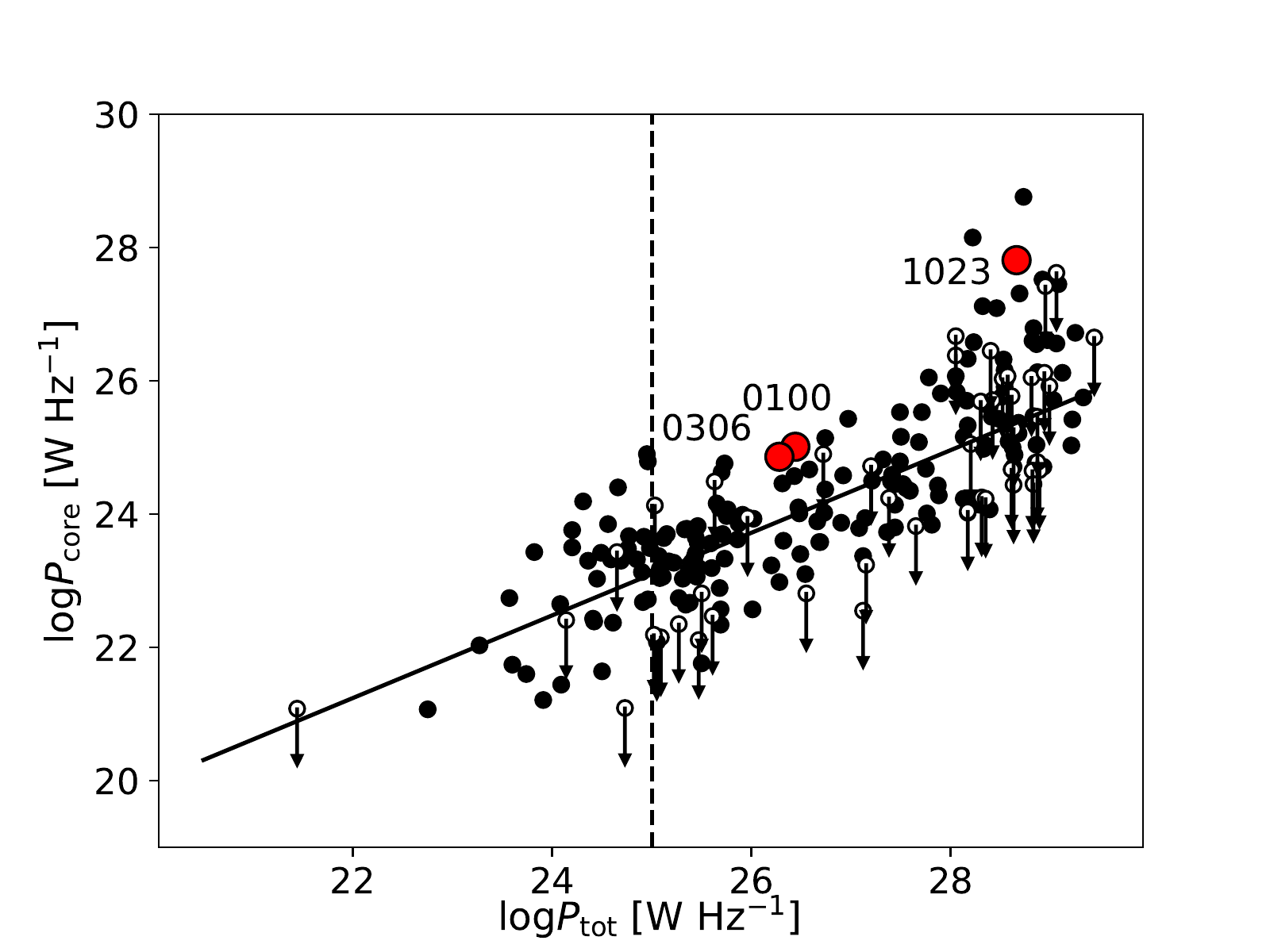}
   \caption{Comparison of core power at 5 GHz and total power at 408 MHz of radio galaxies from 
            \cite{giovannini88,giovannini01,giovannini04}, 
            including all 3C and B2 radio galaxies and radio quasars (black dots), 
            along with the three sources studied here, as labelled (red dots). 
            Because of the high $z$, K-correction was applied to these sources for a comparison with low-redshift sources. 
            The solid black line is the best fit of 3C+B2 sample, while the vertical 
            dashed line highlights the phenomenological division between FRI and FRII 
            sources. According to this division, \jsix\ and \jfive\ can be classified as low-power FRII, 
            while \b2\ is a highly powerful FRII and is distant from the correlation, 
            as expected for a strongly beamed blazar. 
            }
    \label{fig:p_core_tot}%
    \end{figure}

To estimate the source orientation and jet velocity for \jsix\  and \jfive,\ we used the general correlation between the core and total radio power in radio galaxies found by \cite{giovannini88,giovannini01}. 
Since the total radio power was measured at low frequency and is therefore not affected by 
Doppler boosting, the best-fit correlation returns a reference value for the core emission that, 
assuming that sources are oriented at random angles, corresponds to an 
average orientation angle of $\theta = 60^\circ$; the observed dispersion of the core radio power around 
the best-fit line then reflects the different orientation angles \citep[see][for a more detailed discussion]{giovannini04}. 
We can use this correlation to derive the expected intrinsic core radio power from the total galaxy radio power and, comparing it with the observed core radio power, estimate the source orientation.
Figure \ref{fig:p_core_tot} shows 
the core radio power at 5 GHz ($P_{\rm core}$) versus\ the total radio 
power at 408 MHz ($P_{\rm tot}$)
of all 3C and B2 radio galaxies \citep[see][]{giovannini88,giovannini01,giovannini04}, along with our three sources. The radio power at 408 MHz for J0100+2802 was derived from the flux density of the diffuse emission at
1.5 GHz and a spectral index of 1.18. For J0306+1853 we extrapolated the 1.5 GHz flux density with a spectral index of 0.87 and for B2 1023+25 with a spectral index of 0.5. Because of the high redshift, K-correction was applied to the low frequency and core radio power in Fig. 2 for a proper comparison.

The continuous line is the best-fit correlation obtained by \cite{giovannini01}, who 
also properly accounted for the upper limits of the radio core emission.
We also report the average division line between FRI and FRII radio galaxies, namely 
$P_{\rm tot}\sim 10^{25}$ W Hz$^{-1}$ at 408 MHz. 

We can see that B2 1023+25 is among the most powerful sources of the sample. J0100+2802 and J0306+1853 are among the FRII sources, but they are relatively near the separation line.

 All three sources are clearly above the correlation line, and we remember that sources lying above this line are believed to have a strong beaming in their core emission, more intense as 
 the distance from the correlation line increases. 
In fact, \b2, being a known blazar, lies about three orders of magnitude above it.
Both \jsix\ and \jfive\ exceed the expected core power of the correlation, hinting towards a beamed jet, but they are not as extreme as \b2. 
As a cautionary note, we remark that here we are assuming that the correlation between the total and core radio power found at relatively low redshift is still present in sources at high z ($\sim 6$), but 
source evolution and the dominant IC losses suggest that it could evolve.

\jsix\ is core-dominated and, according to the correlation previously discussed \citep[see][for a more detailed discussion]{giovannini04}, we can estimate that the largest possible orientation angle is $\sim 32^\circ$ if the jet velocity is $\Gamma\sim10$. 
The extended emission is one-sided, and the observed asymmetry implies a jet velocity of $\sim0.2$c with the source oriented at $30^\circ$, as expected for a steep spectrum lobe emission. 
At lower orientation angles, the 
required  jet bulk Lorentz factor
strongly decreases; for example,
we have $\Gamma\sim3$ at $\theta\sim13^\circ$ and $\Gamma\sim1.8$ at $\theta\sim3^\circ$.  

\jfive{} is unresolved in our images, and VLBI data
are not available; however, we also applied the core dominance
to this source to constrain its orientation angle, albeit with larger uncertainties.
The largest possible orientation angle is $\sim33^\circ$ with  $\Gamma\sim10$. 
At a smaller orientation angle the required 
jet bulk Lorentz factor
decreases: $\Gamma\sim1.7$ at $\theta\sim3^\circ$ and $\sim1.9$ at $\theta\sim13^\circ$.

For the source B2 1025+23, we used the jet orientation and velocity estimated by \cite{frey15} using VLBI data. 
We note that we obtained similar results using JVLA data and the core dominance method. 

\jsix\ and \jfive{}  have properties similar to low-redshift giant radio galaxies at the centre of rich clusters (cD galaxies). They have a high mass and a radio power in the lower range of FRII sources (see Fig. \ref{fig:p_core_tot}). 
We note that, according to the bivariate radio luminosity function 
\citep[see][]{auriemma77,best05}, the probability of being radio-loud is correlated to the SMBH and galaxy mass, but no correlation is present with the radio power.
The relatively low intrinsic radio power, the moderate jet orientation angle, and the high redshift 
result in a low brightness emission. 
We only detected these sources because we have radio images at microJansky levels. 

The source B2~1023+25 is a high-power blazar. 
Its intrinsic radio luminosity is high, in the range of high-power nearby FRII sources. 
Its jet orientation angle is small, and it appears as a very bright source. 
A similar source at a slightly larger orientation angle is PSO J0309+27 \citep{spingola20}.

\section{Accretion}
\label{sec:accr}

\begin{figure*}
   \centering
   \hspace{-0.5cm}
   \includegraphics[width=\hsize/2]{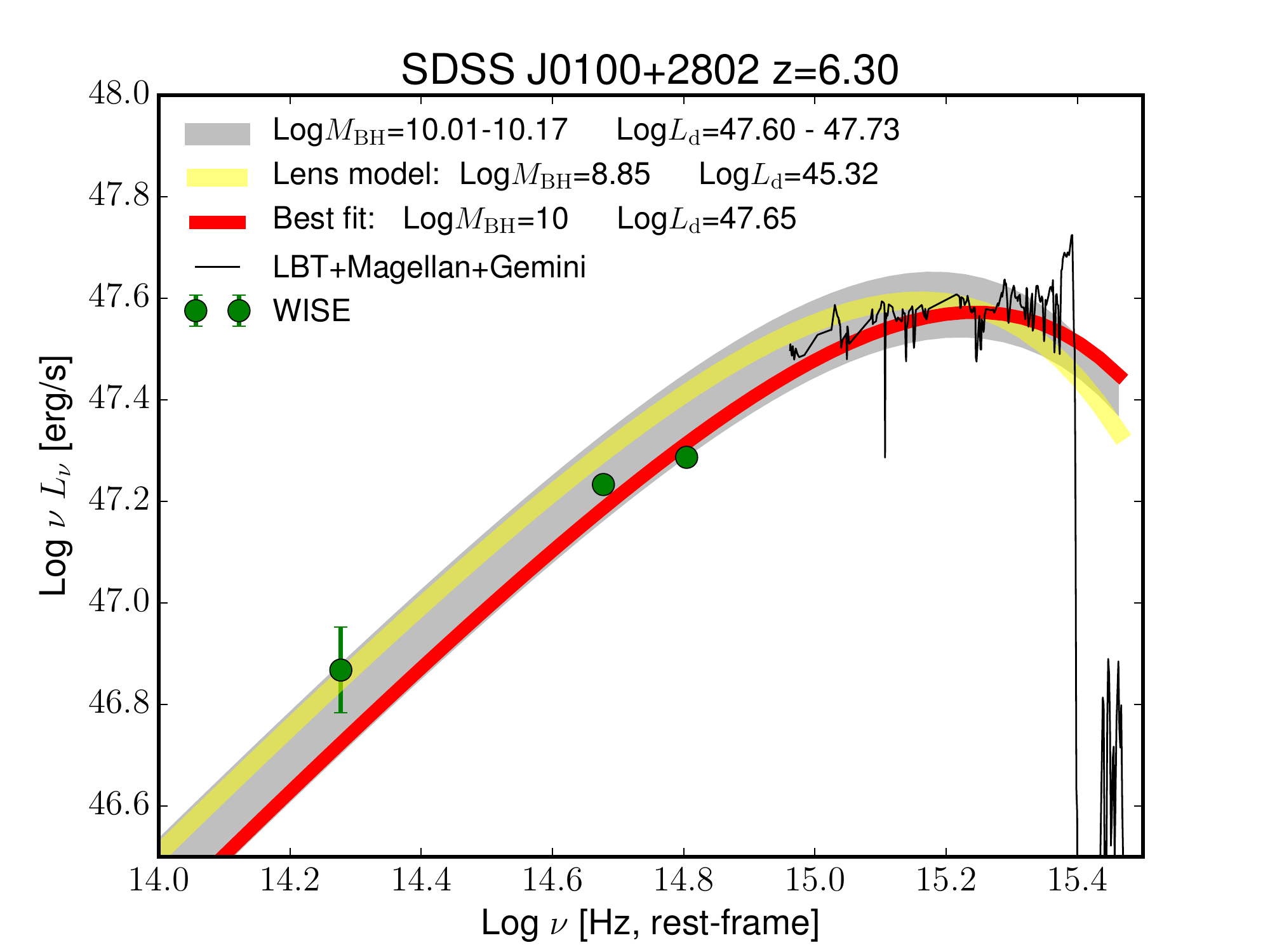}
   \hspace{-0.8cm}
   \includegraphics[width=\hsize/2]{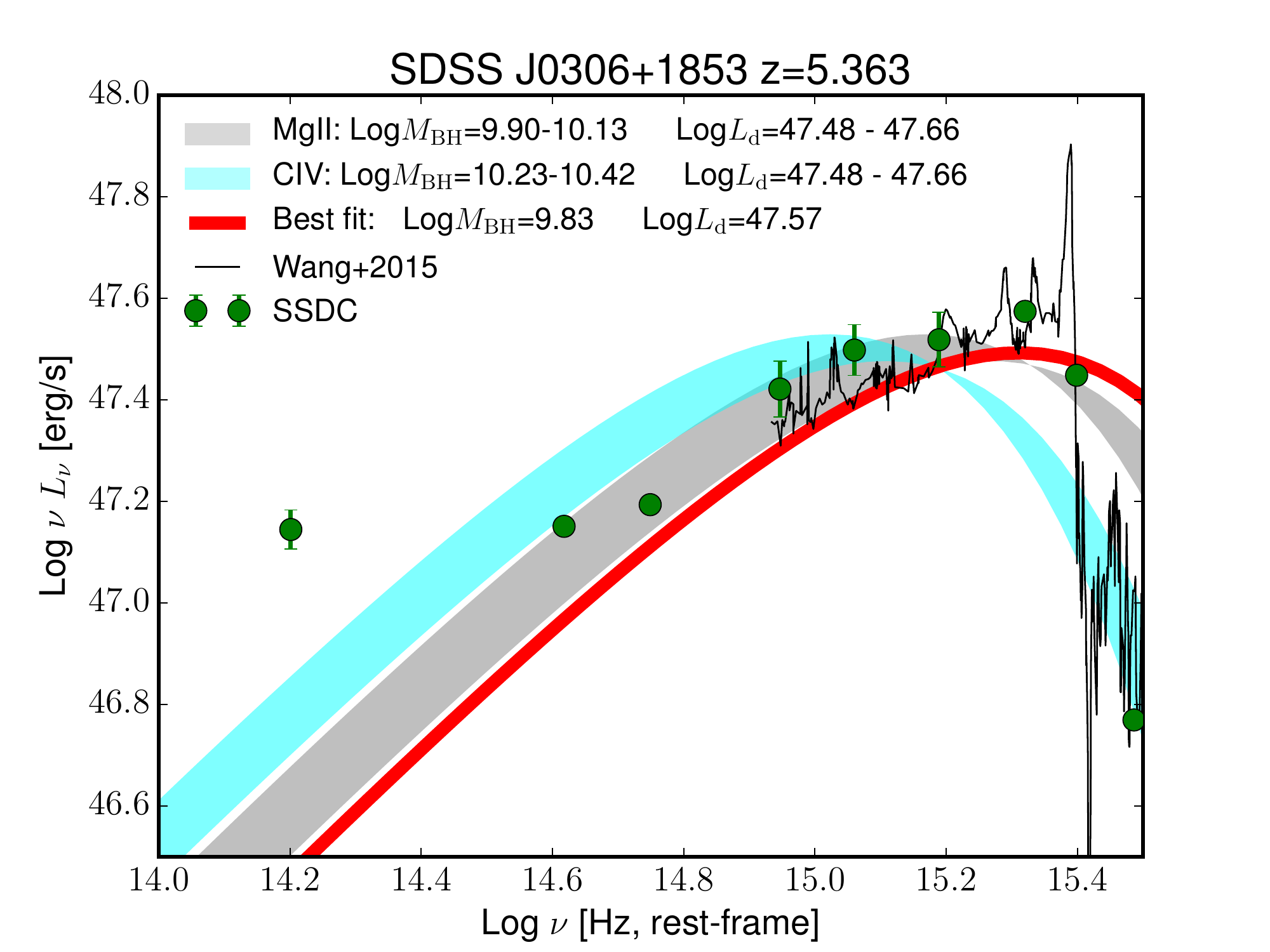}
   \caption{IR-optical-UV SEDs of \jsix{} and \jfive{} (left and right panels, respectively). 
                The solid black line shows the spectrum obtained 
                by \citet[][left]{wu15} and \citet[][right]{wang15}.
                At frequencies larger than Log$\nu$=15.4, 
                the absorption due to intervening Ly$\alpha$ clouds 
                is clearly visible. Green points are {\it WISE} data. 
                The red lines show the best accretion disc fittings. These
                are performed with a \citet{kubota19} model with parameters 
                as labelled and reported in Table \ref{tab:mass}, 
                but it is used to locate the peak of the emission and 
                derive the KERRBB-based mass and spin estimates. 
                The grey band highlights Kubota \& Done profiles obtained with 
                the virial masses and disc luminosities from the 
                bolometric luminosities as derived in \citet[][left]{wu15} and \citet[][right]{wang15}. 
                The cyan shaded region in the right panel shows the same model 
                but built with CIV-based virial masses.
                The yellow line in the left panel shows the same model obtained with the parameters 
                as derived in \citet{fujimoto20}, multiplied by the magnification factor 450.
                }
    \label{fig:disk}%
    \end{figure*}

\begin{figure}
   \centering
   \includegraphics[width=\hsize]{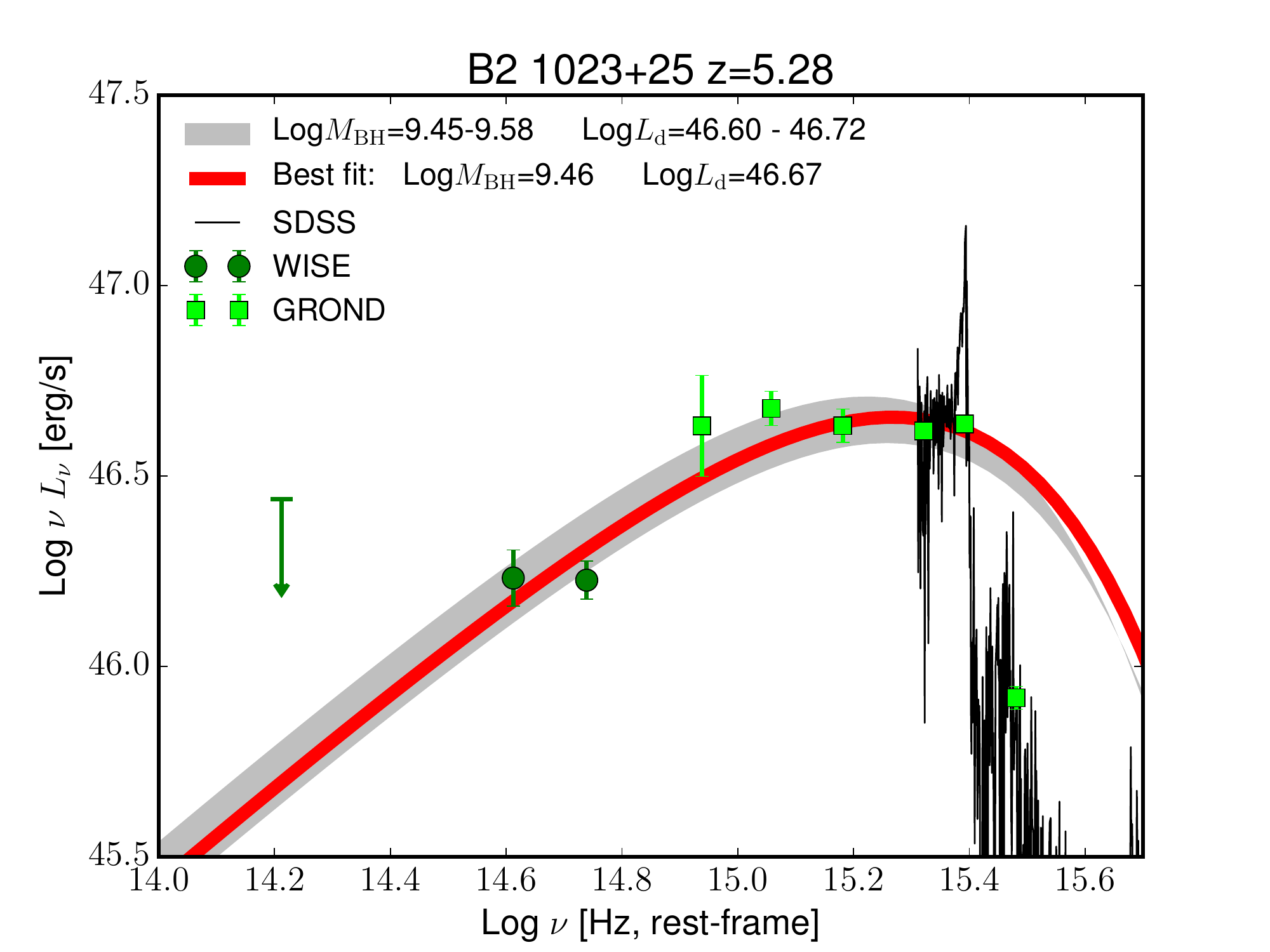}
   \caption{IR-optical-UV SED of B2 1023+25. 
                The solid black line shows the SDSS spectrum, with 
                the clearly visible Ly$\alpha$ clouds absorption at Log$\nu$>15.4.
                Dark green circles are {\it WISE} data, and light green squares are GROND data \citep[]{sbarrato13a}. 
                The red line shows the best accretion disc fitting, while the 
                grey stripe shows the range outside of which the modelling 
                cannot be considered valid, all performed 
                with the \citet{kubota19} model with parameters 
                as labelled.  
                The best fit is used to locate the peak of the emission and 
                derive the KERRBB-based mass and spin estimates. 
        }
    \label{fig:disk-1023}%
    \end{figure}

The black hole masses of \jsix\ and \jfive, estimated through virial-based methods, 
are extremely large, considering that the time elapsed from the Big Bang is 
only $\sim$860 million years for \jsix\ and $\sim$1 billion years for \jfive. 
On the other hand, the uncertainties associated with the `virial methods' are large, 
also because there is no high-redshift calibration to extend their application.
It would be interesting to exploit alternative methods in order to test these values.
To this aim, we  studied the accretion regime and features of all three sources by comparing their optical--near-IR continuum 
emission with two accretion disc models. 
The comparison is shown in Figs. \ref{fig:disk} and \ref{fig:disk-1023}.

%
%

\begin{table}
\caption{Mass estimates, disc and Eddington luminosities (in logarithm), and Eddington ratios
         of \jsix{} and \jfive{} derived according to the methods listed. 
         KD19 refers to the disc fitting method based on \cite{kubota19}. 
         SED Ai+17 refers to the broadband SED fitting results obtained by \cite{ai17}.
         }           
\label{tab:mass}      
\centering          
\begin{tabular}{l c c c l}     
\hline\hline       
\noalign{\smallskip}
Log$M$     & Log$L_{\rm d}$ & Log$L_{\rm Edd}$ & $L_{\rm d}/L_{\rm Edd}$ & Method\\ 
~[M$_\odot$] &  [erg/s]                 &  [erg/s]         & \\  
\noalign{\smallskip}
\hline    
\noalign{\smallskip}
\multicolumn{2}{l}{\it \jsix{}} & & & \\
\noalign{\smallskip}
10.09 & 48.21 & 48.21 & 1    & MgII, $L_{\rm bol}$\\  
      & 47.67 & 48.21 & 0.29 & MgII, $L_{\rm d}$ \\ 
10.00 & 47.65 & 48.11 & 0.34 & KD19\\ 
9.95  & 47.57 & 48.06 & 0.32 & SED Ai+17 \\ 
\hline 
\noalign{\smallskip}
\multicolumn{2}{l}{\it \jfive{}} & & & \\
\noalign{\smallskip}
10.03    & 48.12 & 48.14 & 0.95 & MgII, $L_{\rm bol}$\\  
         & 47.58 & 48.14 & 0.28 & MgII, $L_{\rm d}$ \\ 
10.33    & 48.12 & 48.44 & 0.48 & CIV, $L_{\rm bol}$\\  
         & 47.58 & 48.44 & 0.14 & CIV, $L_{\rm d}$ \\ 
9.83     & 47.57 & 47.94 & 0.43 & KD19\\ 
\hline
\end{tabular}
\end{table}
%

According to 
the virial estimates \citep{wang15,wu15}, the two sources accrete 
close to the Eddington limit, defined as 
$L_{\rm Edd} = 1.26\times10^{38}(M/M_\odot)\,{\rm erg/s}$ 
(see Table \ref{tab:mass}). 
Thus, we first used a modified version of the
geometrically thin, optically thick accretion $\alpha$-disc model 
that better describes the regime close to and 
even beyond the Eddington limit.
This was done through an analytic approximation as described in \cite{kubota19}.
These authors suggest a smooth transition between the standard $\alpha$-disc and an Eddington-limited emission, beginning from the innermost disc region. 
Close to the Eddington limit, in fact, the photon diffusion time across the disc height can exceed the accretion timescale \citep{begelman78}, leading to photon trapping, that is, photons generated in the disc during the viscous process are advected towards the central engine instead of being radiated away.
The change in photon diffusion first starts closer to the central SMBHs due to the higher temperatures reached in those regions. 
\cite{kubota19} follow the assumption that in the photon trapping region the accretion flow 
at a distance $R$ from the black hole
manages to radiate up to the local Eddington flux (defined as 
$F_{\rm Edd}=L_{\rm Edd}/4\pi R^2$),
while the rest of the radiation is advected and ultimately accreted onto the black hole. 
In order to approximate this behaviour, we modified a multi-colour black body by assuming that the temperature is fixed at the equivalent Eddington temperature, 
$T_{\rm Edd}=\left( F_{\rm Edd}/\sigma \right)^{0.25}$ 
($\sigma$ is the Maxwell-Boltzmann constant),
at those radii where the standard \citet{shakura73} temperature would be larger 
than $T_{\rm Edd}$.
This approach allowed us to keep only two free parameters in our modelling: the black hole mass, $M$, and the accretion rate, $\dot M$. 
In principle, the solution also depends on the viewing angle, but we decided for \jsix\ and \jfive\
to fix it at the average value, $\theta_{\rm v}=30^\circ$ (i.e.\ an average value 
between face-on and the expected opening angle of the obscuring torus, $45^\circ$).
B2~1023+25, instead, is a well-known blazar, with a jet viewing angle of $3^\circ$, which we assume to be valid for the accretion disc as well.

It should be noted that we did not perform a proper fitting procedure: We just compared the 
accretion model with photometric and spectroscopic data.
As is clearly visible from Figs. \ref{fig:disk} and \ref{fig:disk-1023},
data at frequencies higher than the Ly$\alpha$ line [Log $(\nu/{\rm Hz})>15.4$ rest frame] 
were not considered, since the flux is absorbed due to intervening clouds of matter 
distributed randomly. 
We did not assume any absorption model, due to their large uncertainties.
Far-IR data are taken in lesser consideration because of the possible contribution from 
the dusty absorbing structure. 
Results of the parameter derivation are listed in Table \ref{tab:mass}: Both \jfive\ 
and \jsix\ require black hole masses of $\sim10^{10}M_\odot$, which is consistent with 
the virial-based masses derived in literature. 
The \b2\ mass ranges between 2.8 and $3.8\times10^9M_\odot$.

The comparison between data and the \cite{kubota19} model is good for all sources, and results are 
consistent with the virial-based ones found in the literature (see Table \ref{tab:mass}), 
but the model relies on the assumption of a non-rotating black hole \citep{kubota19}.
This might be in contrast with the new observations we have presented in the radio band. 
The sources show, in fact, the presence of relativistic jets, which are in general associated 
with highly spinning black holes. 
Thus, in order to include the black hole spin, $a$, 
we considered an analytic approximation 
of the KERRBB model \citep{li05}, initially developed for stellar mass black holes in 
galactic binaries but extended to SMBHs by \cite{campitiello18,campitiello19}.

The data of all three sources show the presence of the peak of an accretion disc 
emission: This is crucial for the approach we are following with KERRBB. 
This model is indeed able to reproduce an $\alpha$-disc-like profile with a 
set of parameters $M$, $\dot M$, $a$, and $\theta$. 
The solutions suffer from strong degeneracy: One should constrain at least two of 
these parameters in order to have a reasonable estimate of the set. 
We first fixed the viewing angle as in our first approach: 
for \b2\ at the known $\theta_{\rm v}=3^\circ$, and in the case of \jsix\ and \jfive\ 
to an average value of $\sim30^\circ$, 
since the sources clearly show their thermal emission and broad emission lines 
(i.e.\ the viewing angle is smaller than the expected torus opening angle, 
$\theta_{\rm T}\sim 40^\circ-45^\circ$). 
As shown in \cite{campitiello18}, the remaining parameters are degenerate, 
and the same peak position can be reached by accretion discs with different 
$M,\dot M,$ and $a$ values, between $a=-0.998$ and $a=0.998$\footnote{
The maximally co- and counter-rotating spin values $a=+1, -1$ cannot be 
reached if capture of radiation by the black hole is considered \citep{thorne74}.
}.
Table \ref{tab:kerr} shows the results obtained by following \cite{campitiello18,campitiello19} 
on the peak positions derived with the \cite{kubota19} approach, for the notable 
spin values of $a=0,0.998,-0.998$. 
All results point towards extreme black hole mass values and high Eddington ratios that span between $L_{\rm d}/L_{\rm Edd}=0.34$ and the Eddington limit. 
Since \jfive\ and \jsix\ present jet signatures, their central black holes are 
expected to be maximally spinning and co-rotating with the accretion disc. 
This favours larger black hole masses and smaller Eddington ratios.
We discuss the implications of these expectations in Sect. \ref{sec:disc}.

%
\begin{table}
\caption{KERRBB results derived  from the best-fit peak position. 
         The columns are the black hole spin, the inner radius in units of the gravitational radius, the logarithmic values of the black hole mass (per solar mass), the accretion rate (in units of g/s), the disc luminosity in units of erg/s, and the Eddington ratio defined in terms of luminosities.
         }             
\label{tab:kerr}      
\centering          
\begin{tabular}{c c c c c c c }     
\hline\hline       
\noalign{\smallskip}
$a$ & $R_{\rm in}/R_{\rm g}$ & ${\rm Log}\frac{M}{M_\odot}$ & ${\rm Log}\frac{\dot M}{\rm g/s}$ & ${\rm Log}\frac{L_{\rm d}}{\rm erg/s}$ & $L_{\rm d}/L_{\rm Edd}$\\ 
\noalign{\smallskip}
\hline                    
\noalign{\smallskip}
\multicolumn{3}{l}{\it \jsix{}} & & & \\
\noalign{\smallskip}
0           & 6     & 9.78  & 28.00 & 47.71 & 0.66\\  
0.998           & 1.24  & 10.24 & 27.48 & 47.94 & 0.39\\ 
--0.998         & 9     & 9.62  & 28.16 & 47.69 & 0.90\\ 
\hline                    
\noalign{\smallskip}
\multicolumn{3}{l}{\it \jfive{}} & & & \\
\noalign{\smallskip}
0           & 6     & 9.64  & 27.92 & 47.63 & 0.75\\ 
0.998       & 1.24  & 10.10 & 27.40 & 47.86 & 0.44\\ 
--0.998         & 9     & 9.48  & 28.08 & 47.61 & 1.03\\ 
\hline           
\noalign{\smallskip}
\multicolumn{3}{l}{\it B2 1023+25} & & & \\
\noalign{\smallskip}
0           & 6     & 9.26  & 27.04 & 46.75 & 0.24\\ 
0.998       & 1.24  & 9.66  & 26.56 & 47.02 & 0.18\\ 
--0.998         & 9     & 9.11  & 27.19 & 46.73 & 0.32\\ 
\hline                  
\end{tabular}
\end{table}


\section{Jet features}
\label{sec:jet}

All three sources clearly show the presence of jets in their radio emission. 
B2 1023+25 has already been classified as a blazar because of its radio-loudness, 
flat radio spectrum, strong and hard X-ray flux and spectrum, and super-luminal motion 
in VLBI images \citep[]{frey15}. 

\jsix{} and \jfive{}, instead, are defined as radio-quiet sources: They are in the 
FIRST survey footprint, without being detected. 
\jfive\ was never detected in the radio band, leading to a radio-silent classification. 
At the microJansky level, on the other hand, these sources are well detected, showing that 
they are indeed powerful in the radio band and host a jet. 
We calculated their radio-loudnesses, defined as 
$R_{\rm L}=F_{\rm 5\, GHz}/F_{\rm 4400\, \AA}$ rest frame, 
in order to compare them with the known quasar population. 
Even if their radio power is intense (cf.\ Table \ref{tab:radio-obs}), their 
extremely luminous big blue bumps led us to classify them as radio-quiet sources, 
with $R_{\rm L} = 0.45$ for \jsix{} and $R_{\rm L} = 0.7$ for \jfive. 

We note that \cite{padovani16} already proposed dividing AGNs not into radio-loud and
radio-quiet categories, but into jetted and non-jetted, where `jet' here means 
a relativistic jet. 
This is certainly a more physical classification and is even more appropriate
when dealing with high-redshift sources that have a very strong accretion disc
and whose extended (quasi-isotropic) radio flux could be quenched by the 
enhanced inverse Compton cooling off the CMB photons.

Their radio features are not very common in such strongly accreting 
quasars: Locally, one would expect to see strong jets, culminating in extended 
emission, in sources with such high accretion power \citep{ghisellini14b,ghisellini15}.
\jsix{} and especially \jfive{}, instead, show rather compact emissions, with no signs 
of strong extended lobe-like structures. 
Only \jsix{} shows a jetted structure, resolved 
on VLA scale, which allowed us to discriminate between compact and diffuse emissions. 
\jfive\ instead presents a compact and rather steep radio emission: We cannot exclude 
that it hosts a young jet, such as those shown in compact steep spectrum (CSS) 
or gigahertz peaked spectrum (GPS) radio sources \citep{odea98,odea21}. 

In the following we briefly summarize the model we used to fit the jet emission
in order to find its main physical properties, including its total power.
The latter is the main parameter we need to evaluate the power injected into the
extended structure, which can be a hot spot or a lobe.
The distinction between hot spot and lobe, when we lack detailed radio maps,
is somewhat arbitrary: For simplicity, we call a structure with 
a size (radius) of less than 3--5 kpc a `hot spot', and above this size a `lobe'.

We stress that our model for the extended structure assumes a spherical and
homogeneous structure, without any gradient.
When more precise data are missing, such as the level of the high energy
emission, we try to be as close as possible to the case of equipartition 
between the magnetic and the relativistic electron energies.

\subsection{Modelling the inner jet emission}

As mentioned, of our three sources only B2 1023+25 is clearly a blazar, 
and it has a very powerful jet that points towards us. 
To model it, we used the model described in detail in \cite{ghisellini09}.
We recall here only the basic features of the model,
following the same notation as the original paper.

The emitting region is assumed to be a sphere of radius $r$ at a distance $R_\mathrm{diss}$ 
from the black hole, moving 
with a bulk Lorentz factor $\Gamma$ at an angle $\theta_{\rm v}$ from the observer.
The jet is assumed conical with an aperture angle of 0.1 rad. 
A particle distribution $Q(\gamma)$ of relativistic electrons 
of energy $\gamma m_{\rm e} c^2$
is injected 
throughout the region, for a light crossing time $r/c$. 
The injected particle distribution is a smoothly broken power law:
\begin{equation}
Q(\gamma) \, =\, Q_0 \, \frac{ (\gamma/\gamma_{\rm b})^{-{s_1}} }{
 1 + (\gamma/\gamma_{\rm b})^{-{s_1}+{s_2}} }
 \label{qgamma}
,\end{equation}
where $\gamma_{\rm b}$ is the break energy,
$s_1$ and $s_2$ are the $Q(\gamma)$ slopes below and above the break, and $Q_0$ is the distribution normalization.
The particle density distribution, used to calculate the spectrum, is obtained 
through the continuity equation that accounts for injection, radiative losses, 
photon--photon collisions, and pair production.
For the radiative losses, we account for synchrotron, self-Compton and inverse Compton scatterings off the disk, broad lines, torus, and CMB radiation (external Compton).
We also calculate the jet power, that is, the sum of the power emitted and carried 
in magnetic fields and particles.
All these powers can be expressed as 
\begin{equation}
P_i \, =\, 4 \pi r^2 c \Gamma^2 U^\prime_i
,\end{equation}
where the subscript $i$ refers to the different forms of energy and $U^\prime_i$ 
is calculated in the co-moving frame.
The total power is the sum.
In general, the magnetic and particle powers are model-dependent, while the radiative 
power depends only on the $\Gamma$ factor since $U_i = U_{\rm rad}$ is
\begin{equation}
U_{\rm rad} \, \sim \, \frac{L^\prime}{4\pi r^2 c}  = 
\frac{L^{\rm obs}}{4\pi r^2 c \delta^4}
,\end{equation}
where $L^\prime$ is the luminosity measured by the co-moving observer.
When $\theta_{\rm v}\sim 1/\Gamma,$ we have $\delta=\Gamma$ and then:
\begin{equation}
P_{\rm rad} \, \sim \,  \frac{L^{\rm obs}}{\Gamma^2}
.\end{equation}
This can be taken as a very robust lower limit to the jet power.
For our purposes, the jet power is an important parameter because we can use 
it to estimate the power that can energize the extended structures, such as 
hot spots and lobes.

\subsection{Modelling the extended structures}

For the extended structures we followed the model described in 
\cite{ghisellini14a}.
We take into account the radiative and adiabatic cooling of the emitting electrons,
considering the effect of the external Compton (EC) scattering off the CMB radiation.
Its energy density, $U_{\rm CMB}$, scales as $(1+z)^4$, and this implies that the 
corresponding EC luminosity at high energies becomes dominant over SSC when
the magnetic field energy density is smaller than $U_{\rm CMB}$, giving:
\begin{equation}
B_{\rm CMB} \, = \, [8\pi U_{\rm CMB}]^{1/2} = 3.26\times 10^{-6} (1+z)^2 \,\, {\rm G}
.\end{equation}
At $z=6,$ we have $B_{\rm CMB} = 160$ $\mu$G.
This has two effects: First, the enhanced EC cooling implies a strong X-ray emission,
and, second, the enhanced cooling implies a steepening of the particle distribution 
at high energies, with a corresponding dimming of the high frequency 
radio emission. 
We stress that this can only be found by using the continuity equation,
while the assumption of a given, fixed particle distribution would not show
the `radio quenching' effect. 

The adiabatic cooling rate is:
\begin{equation}
 \dot\gamma_{\rm ad} \, =\, \frac{ \gamma \beta_{\rm exp}}{r/c}
,\end{equation}
where $\beta_{\rm exp} c$ is the expansion speed of the lobe or hot spot. 
As for the injected particle distribution, we used the same function as
in Eq. \ref{qgamma}, but with different slopes and $\gamma_{\rm b}$ values.
The total power injected, in relativistic electrons, 
throughout the extended structure is assumed
to be a fraction, $\epsilon$, of the jet power, namely 
\begin{equation}
V\, \int Q(\gamma) \, \gamma \, m_{\rm e} c^2 \, d(\gamma) \, =\, \epsilon P_{\rm jet}
,\end{equation}
where $V$ is the source volume, assumed spherical for the sake of simplicity. 

We then calculated the total magnetic and particle energy in the extended structure, making use of the particle energy distribution $N(\gamma)$ 
found using the continuity equation and neglecting protons, assumed 
to be cold:
\begin{equation}
E_{\rm e}=V \, \int N(\gamma) \gamma m_{\rm e} c^2 d\gamma; 
\quad  E_B = V \frac{B^2}{8\pi }
.\end{equation}
Equipartition then corresponds to $E_{\rm e}=E_B$.

%
%
 \begin{figure*}
   \centering
   \vskip -2.5 cm
   \hspace{-0.17cm}
   \includegraphics[width=\hsize/2]{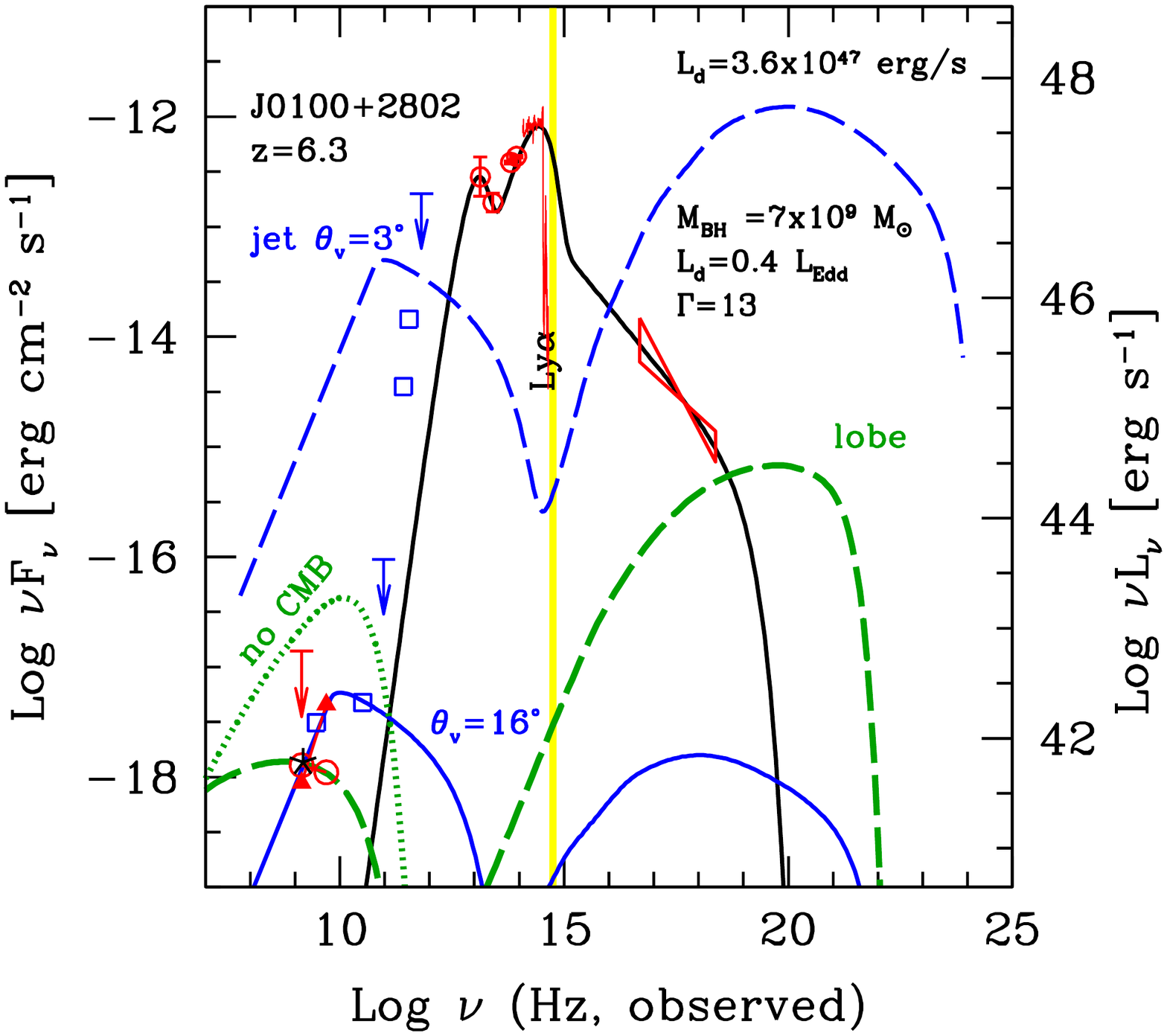}
   \hspace{-0.17cm}
   \includegraphics[width=\hsize/2]{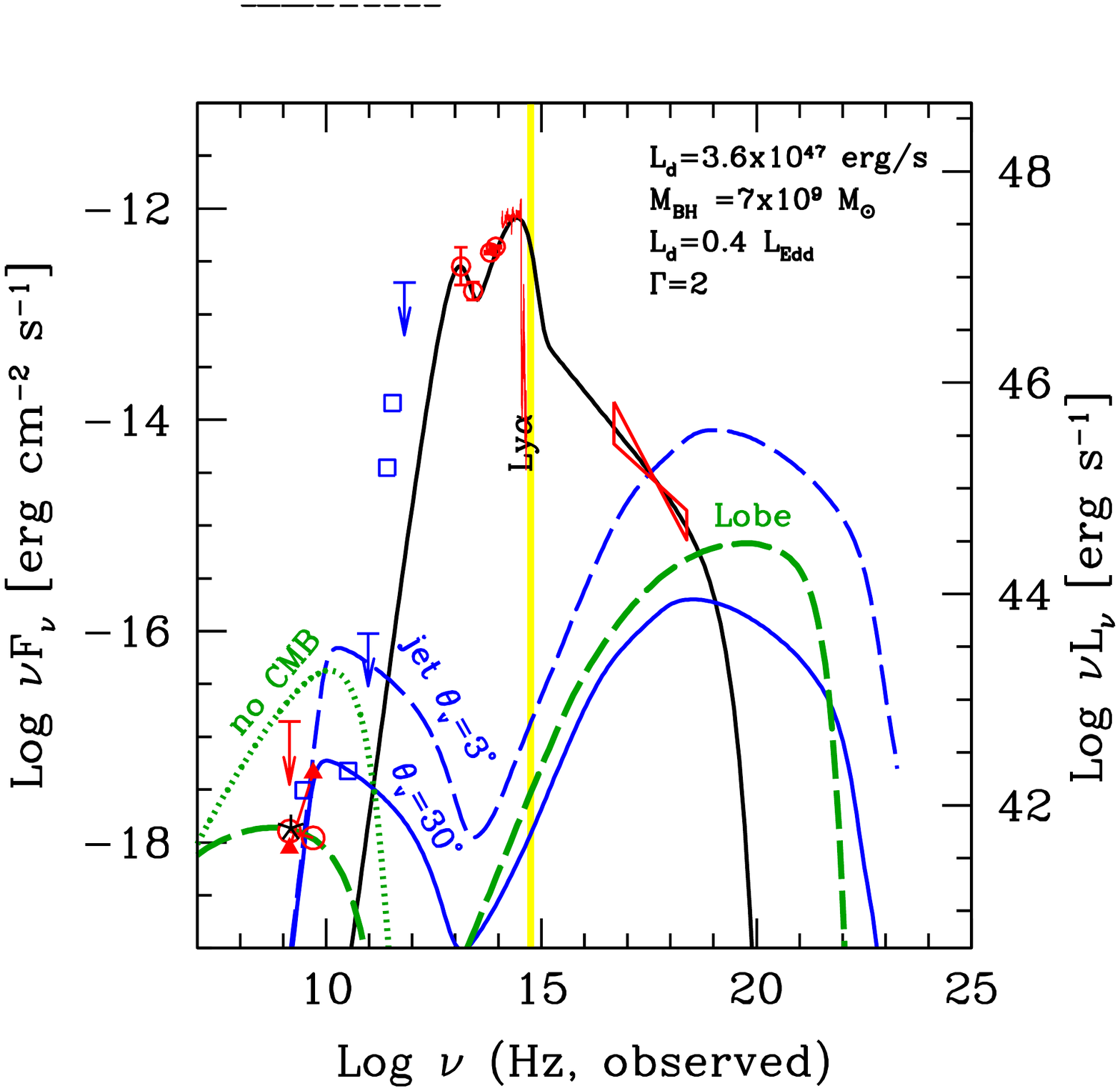}
   \vskip -2.5 cm
   \caption{
   \jsix\ SED assembled from this work, archival data (Space Science Data Center), \cite{wang16}, and \cite{ai17}. In the radio band, the VLA data from this paper are shown as triangles (from the core) and empty circles (from the extended structure). The solid black  line corresponds to the accretion disc, the X-ray corona, and the molecular torus. {\it Left panel: }
  Emission of a powerful jet with $\Gamma=13$ (dashed blue  line) as would be seen if the viewing angle were $\theta_{\rm v}=3^\circ$. The solid blue line is the same jet observed at $\theta_{\rm v}=16^\circ$, to fit the core VLA emission. The green dashed line corresponds to the emission of a lobe of radius of 70 kps. 
  {\it Right panel: } Emission of a jet with $\Gamma=2$ (dashed blue line) as would be seen if the viewing angle were $\theta_{\rm v}=3^\circ$. The solid blue line is the same jet observed at $\theta_{\rm v}=30^\circ$, to
  fit the core VLA emission. The dashed green line corresponds to the emission of the same lobe as in the left panel. The set of parameters is in Tables \ref{parajet} and \ref{paralobe}.
}
    \label{0100}
    \end{figure*}
 \begin{figure*}
   \centering
   \hspace{-0.17cm}
   \includegraphics[width=\hsize/2]{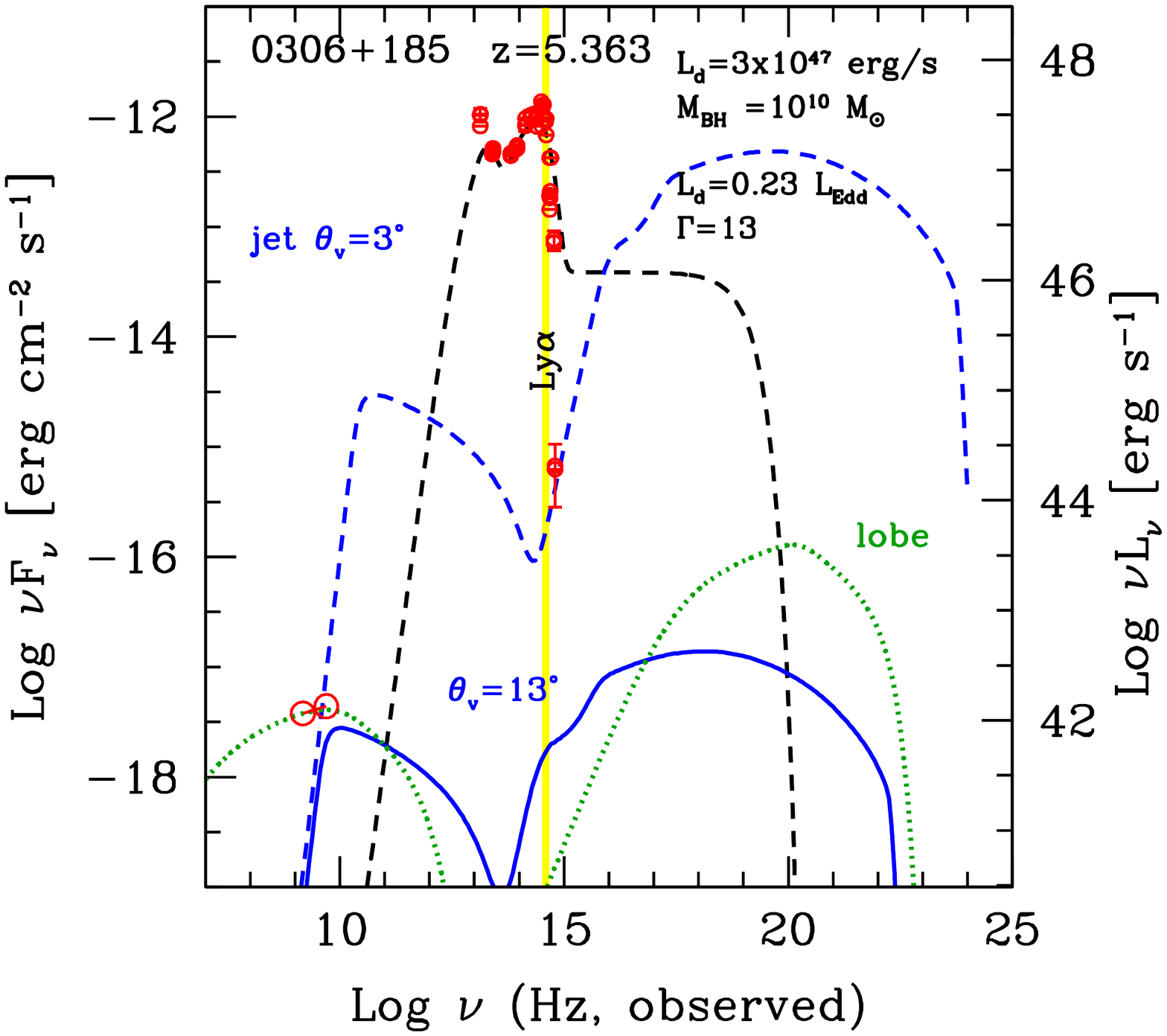}
   \hspace{-0.17cm}
   \includegraphics[width=9cm]{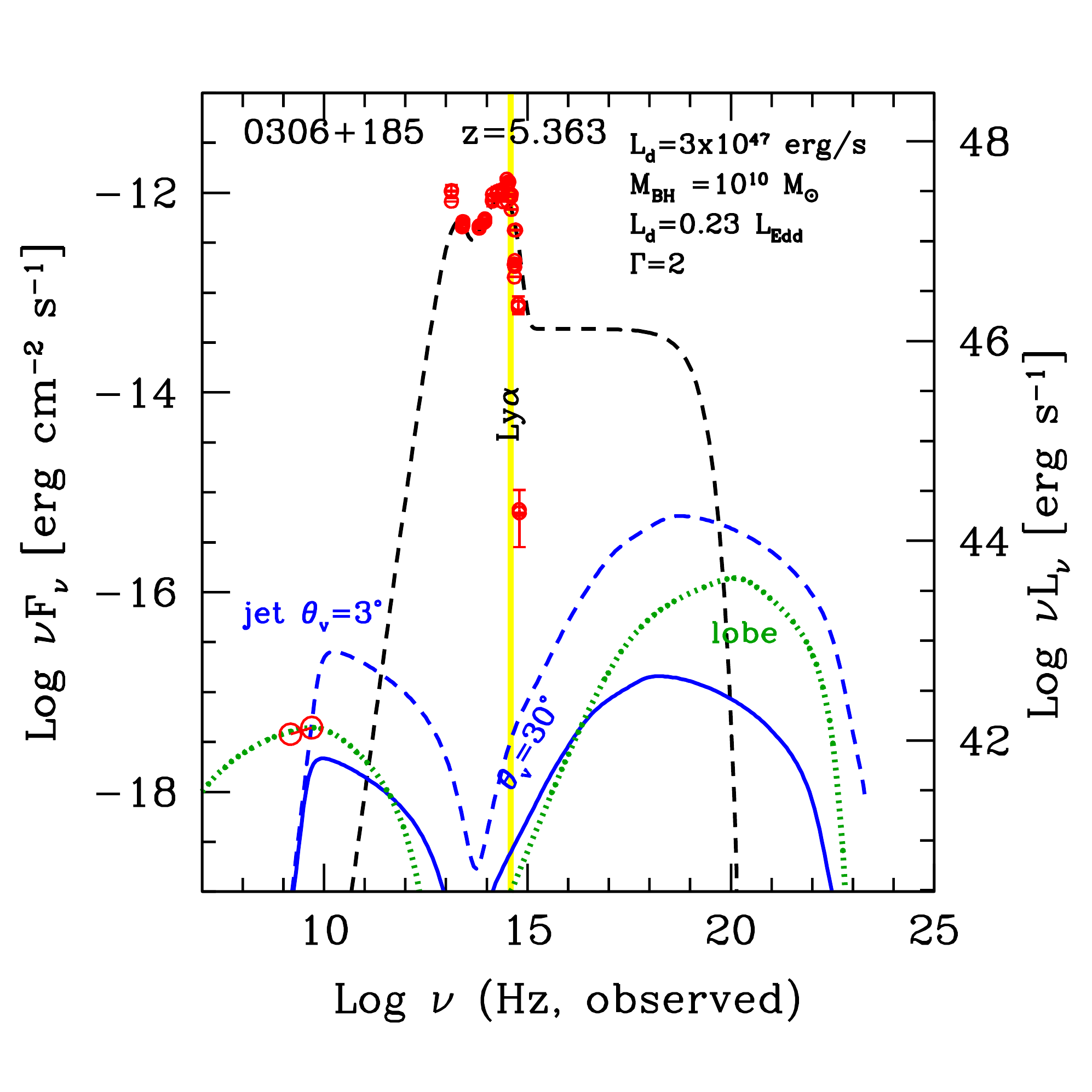}
   \vskip -0.5 cm
   \caption{SED of \jfive\ (archival data from SSDC). The empty circles in the radio band are our VLA data at 1.4 and 5 GHz. The corresponding spectral index is $\alpha = 0.88$ ($F_\nu \propto \nu^{-\alpha}$). The solid black line corresponds to the accretion disc, the X-ray corona, and the molecular torus. {\it Left panel: }  Emission from a powerful jet with $\Gamma=13$ (dashed blue line), as would be seen with a viewing angle of $3^\circ$. 
If we interpret the VLA data as the sum of a flat core plus a steep extended component, then both structures should substantially contribute to the flux.
This requires the jet to be seen misaligned with $\theta_{\rm v}=13^\circ$
(solid blue line). The extended structure is assumed to be a lobe with a radius of 10 kpc, in equipartition (namely $E_{\rm B}=E_{\rm e}$; dotted green line). {\it Right panel:} Emission from a weaker jet with $\Gamma=2$ (dashed blue line), as seen with $\theta_{\rm v}=3^\circ$. To contribute to the radio VLA data, $\theta_{\rm v}=30^\circ$ is required. The extended structure (dotted green line) is the same as in the left panel. The set of parameters is listed in Tables \ref{parajet} and \ref{paralobe}.
}
    \label{0306}
    \end{figure*}
 \begin{figure}
\vskip -2.5 cm
   \centering
   \includegraphics[width=\hsize]{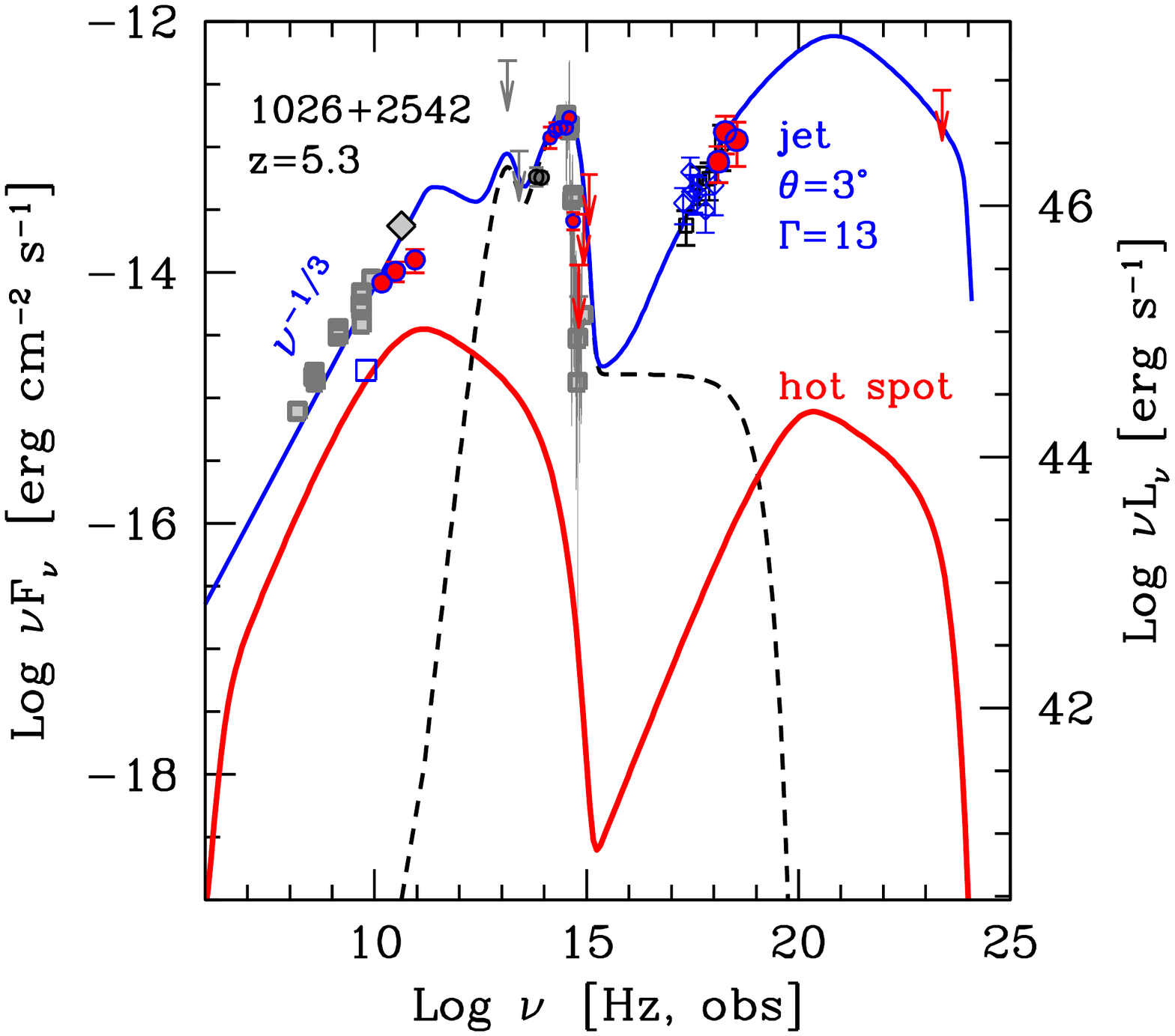}
\vskip -2.5 cm
   \caption{SED of \b2, clearly showing its blazar nature. We do not have information on the extended structure, besides a flux of 32 mJy at 1.4 GHz, shown by the empty blue square in the radio band. 
   The lobe is therefore only representative of a structure of 50 kpc in size, reprocessing 10\% of the jet power, with magnetic field and relativistic electrons in equipartition. The set of parameters is listed in Tables $\ref{parajet}$ and \ref{paralobe}.
                }
    \label{1026}
    \end{figure}
\begin{table*} 
\centering
\begin{tabular}{l l l l l l l l l l l l l l l l l l}
\hline
\hline
Source &$z$ &$M$ &$L_{\rm d}$ &$R_{\rm diss}$ &$R_{\rm BLR}$ &$P^\prime_{\rm e, jet, 45}$  &$B$ &$\Gamma$ &$\theta_{\rm V}$  &$\gamma_{\rm b}$ &$\gamma_{\rm max}$ &$s_1$ &$s_2$   
  &$P_{\rm jet, 45}$ \\ 
~[1] &[2] &[3] &[4] &[5] &[6] &[7] &[8] &[9] &[10] &[11] &[12] &[13] &[14] &[15]  \\ \\
\hline   
0100+2802 &6.3   &7e9  &364  &2519 &1907 &0.03 &1   &13 &16  &70  &3e3 &1 &2.6 &366   \\   
0100+2802 &6.3   &7e9  &364  &1680 &1907 &0.06 &0.4 &2  &30  &70  &3e3 &1 &2.6 &9.4  \\ 
\\
0306+185 &5.363 &1e10  &300  &1800 &1729 &0.01  &0.8 &13 &13 &70  &3e3 &1.5 &2.6 &142  \\
0306+185 &5.363 &1e10  &300  &1800 &1729 &5e--3 &1.2 &2  &30 &70  &3e3 &1.5 &2.6 &1.9  \\ 
\\
\b2\ &5.3  &2.8e9 &58.2 &840  &763  &0.01  &2.6 &13 &3  &90  &4e3 &0   &2.6 &62  \\
\hline
\hline 
\end{tabular}
\vskip 0.4 true cm
\caption{
Adopted parameters for the jet model shown in Figs. \ref{0100}--\ref{1026}.
Column [1]: source; 
Col. [2]: redshift;
Col. [3]: black hole mass in solar masses;
Col. [4]: disc luminosity in units of $10^{45}$ erg s$^{-1}$;
Col. [5]: distance of the dissipation region from the black hole, in units of $10^{15}$ cm;
Col. [6]: size of the broad line region (BLR), in units of $10^{15}$ cm;
Col. [7]: power injected in the jet in relativistic electrons, calculated in the co-moving 
frame, in units of $10^{45}$ erg s$^{-1}$;
Col. [8]: magnetic field in G;  
Col. [9]: bulk Lorentz factor;
Col. [10]: viewing angle in degrees;
Cols. [11] and [12]: break and maximum Lorenz factor of the injected electron distribution;
Cols. [13] and [14]: slopes of the injected electron distribution; 
Col. [15]: Total jet power, in units of $10^{45}$ erg s$^{-1}$.
The values of the powers and the energetics refer to one jet.
}
\label{parajet}
\end{table*}
\begin{table*} 
\centering
\begin{tabular}{l l l l l l l l l l l l l l  l l l}
\hline
\hline
Source &Comp. &$R$  &$P_{\rm e,45}$ &$B$  &$\gamma_{\rm b}$ &$\gamma_{\rm max}$ 
&$s_1$ &$s_2$ 
&$\log E_{\rm e}$  &$\log E_{\rm B}$ &$\beta$ \\
~[1] &[2] &[3] &[4] &[5] &[6] &[7] &[8]&[9] &[10] &[11] &[12] \\
\hline   
0100+2802 &lobe &70 &1.1  &7  &4e3  &1e5  &1  &2   &58.1 &58.9 &0.2   \\
0306+185 &lobe &10 &2  &21.7 &1e3  &3e5  &1  &2.7  &57.4 &57.3 &0.14    \\ 
\b2\ &hot spot &3 &10 &250    &1e4  &1e6  &1.3 &2.5 &57.9 &57.4 &0.2 \\
\hline
\hline 
\end{tabular}
\vskip 0.4 true cm
\caption{
Adopted parameters for the hot spot and lobe models shown in Figs. \ref{0100}-\ref{1026}.
Col. [1]: source; 
Col. [2]: component; 
Col. [3]: size in kpc;
Col. [4]: power injected in relativistic electrons in units of $10^{45}$ erg s$^{-1}$;
Col. [5]: magnetic field in $\mu$G;
Cols. [6] and [7]: break and maximum Lorenz factor of the injected electron distribution;
Cols. [8] and [9]: slopes of the injected electron distribution; 
Col. [10]: logarithm of the total energy in relativistic electrons, in erg;
Col. [11]: logarithm of the total energy in the magnetic field, in erg;
Col. [12]: bulk velocity of the lobe or the hot spot.
The values of the powers and the energetics refer to each hot spot and lobe.
}
\label{paralobe}
\end{table*}

\subsection{\jsix: The powerful jet case} 

Figure \ref{0100} shows the SED of \jsix\ together with the models of the jet
and the extended emission. 
The data are from the Space Science Data Center archive (SSDC\footnote{\url{
https://tools.ssdc.asi.it/}}
), from \cite{wang16} and \cite{ai17}, 
while the radio data are from this work.
As discussed before, the accretion disc emission is clearly visible, as is the
X-ray spectrum, which was already analysed and discussed in \cite{ai17} and interpreted as
emission from the X-ray corona. 
From our data, we can see the presence of two radio components, whose fluxes are
shown by the empty circles and filled triangles, respectively.
The triangles suggest a very flat (or inverted) component, which we interpret as
emission from a compact relativistic jet. 
The other component (circles) is instead steep, and we interpret that as coming
from an extended structure, with a size of $\sim$ 70 kpc (projected).

To model the jet of this source we have only the JVLA radio data of the flat component.
It is therefore necessary to use some general guidelines in order to find a meaningful
solution. 
We first assumed that the jet of \jsix\ is a `standard' jet, namely that,
if viewed at a small viewing angle, it should appear as a very powerful blazar jet,
with a power comparable to, if not larger than, the disc luminosity.
On the other hand, when seen misaligned, it must produce the JVLA flux
of the flat component. 
This is shown in the left panel of Fig. \ref{0100}: The dashed blue
line is the jet emission that a $\theta_{\rm v}=3^\circ$ observer would see.
In this case we have comparable disc luminosity and the total jet power,
as listed in Table \ref{parajet}.
The same jet, if observed at $\theta_{\rm v}=16^\circ$, would account for
the flat JVLA component, as shown by the solid blue line. This is in contrast with the discussion presented in Sect. 2.2 and implies that the known correlation between the core and total radio power in low-redshift AGNs needs to be revised at high $z$.

For the extended component (dashed green line) we have the constraint of the observed, 
relatively steep X-ray flux (the extended structure should not overproduce it)
and the fact that in the radio map there is no sign of a counter-lobe.
This suggests that the entire structure is not stationary but rather has some bulk motion.
We find that $\beta\sim 0.2$ is enough to make the counter-jet not detectable.
The solution shown in Fig. \ref{0100} assumes that the injected power, needed
to accelerate the electron in the lobe, is 0.3\% of the total jet power
($1.1\times 10^{45}$ vs. $3.7\times 10^{47}$  erg s$^{-1}$). 
The lobe has $E_{\rm B}\sim 6 E_{\rm e}$.

To show the importance of the CMB cooling, we also show the same model, but setting
$U_{\rm CMB}$ to zero. 
The synchrotron flux would be enhanced, at the expense of the X-ray flux,
which precipitates below the scale of the figure (at the level 
of $10^{-20}$ erg s$^{-1}$ cm$^{-2}$).

\subsection{\jsix: The weak jet case} 

In this case we assumed that the jet is still mildly relativistic,
but with $\Gamma=2$ and $L_{\rm d} \sim 39 P_{\rm jet}$.
The reason for this choice comes from the correlation
discussed in Sect. 2.2 between the core and the total flux of radio AGNs,
shown in Fig. \ref{fig:p_core_tot}. 
In this case, the de-beamed flux is much less dependent on the viewing angle,
allowing it to be greater.
The corresponding model is shown in the right panel of Fig. \ref{0100}. 
The figure shows the appearance of the jet flux if $\theta_{\rm v}=3^\circ$
(dashed blue line) and at $\theta_{\rm v}=30^\circ$ (solid blue line).
We have chosen $\theta_{\rm v}=30^\circ$ because it is the average angle
considering the presence of a molecular absorbing torus with a semi-aperture angle
of $\theta_{\rm torus} \sim 45^\circ$. 
At angles larger than $\theta_{\rm torus}$ the optical flux would be severely absorbed,
and the source would not have the chance to be detected by the SDSS survey.
The lobe model is the same as in the left panel.
From Tables \ref{parajet} and \ref{paralobe}
we see that the extended lobe requires 12\% of the total jet power.

\subsection{\jfive: The powerful jet case} 

The rationale behind the parameter choice for the powerful jet in \jfive\ is the
same as before: The jet has a total power close to $L_{\rm d}$ and $\Gamma=13$.
The left panel of Fig. \ref{0306} shows this case, and the dashed blue line is what
a $\theta_{\rm v}=3^\circ$ observer would see.
The model is weakly constrained by the measured spectral index between 1.5 and 5 GHz,
suggesting the contribution of both a flat (core) and a steep (extended)
component. 
The viewing angle for which the jet emission comes close to the JVLA data is
$\theta_{\rm v} \sim 13^\circ$, and the corresponding jet model is shown by the solid blue line.

For the extended component, from the JVLA radio map we assumed a lobe size of 10 kpc (de-projected), and the absence of a counter-lobe can be explained by a bulk velocity 
of the lobe of $\beta=0.14$.
The power needed for this structure is $2\times 10^{45}$ erg s$^{-1}$, to be compared 
with a total jet power of $P_{\rm jet}\sim 1.4\times 10^{47}$ erg s$^{-1}$ (i.e.\ 71 times larger).
The lobe would dominate the extended X-ray flux, but a sensitivity larger than 
$10^{-16}$ erg s$^{-1}$ cm$^{-2}$, together with an excellent
angular resolution, is required to detect it.

\subsection{\jfive: The weak jet case} 

In complete analogy with the case of \jsix, we again assumed $\Gamma=2$ together
with a total jet power much lower than the disc luminosity (i.e. 150 times smaller).
The resulting model is shown in the right panel of Fig. \ref{0306}:
The dashed blue line corresponds to the $\theta_{\rm v}=3^\circ$ observer, 
and the solid blue line to $\theta_{\rm v} = 30^\circ$.
The model for the extended structure is the same as in the left panel, 
but now the power it requires exceeds the total jet power,
disfavouring this solution.

\subsection{\b2}
For this source there is no ambiguity between a strong and weak jet, since
it is a well-known blazar, as discussed in \cite{sbarrato12,sbarrato13b}.
We updated the SED to include the flux estimated by the JVLA data (i.e.\ 32 mJy at 5 GHz). 
This is the result of a subtraction of 
our JVLA flux and the VLBI flux observed by \cite{frey15}.
We then assumed that this flux is produced by a hot spot with a radius
of 3 kpc, moving with a bulk motion of $\beta\sim 0.2$.
To explain the 32 mJy flux, and  to not be far from equipartition, 
the hot spot requires one-sixth of the total jet power (i.e.\ $10^{46}$ vs. $6\times 10^{46}$
erg s$^{-1}$).
We note that the magnetic field of the hot spot (250 $\mu$G) is larger than $B_{\rm CMB}$
for $z=5.3$, and therefore the emission is not affected by a strong EC cooling.

\subsection{Discussion}

For \b2,\ there is no ambiguity in the power of its jet.
It follows the relation between jet power and accretion disc luminosity 
put forward by, for example, \cite{ghisellini14b}, making \b2\ a powerful blazar.
The other two sources, on the other hand, definitely have a jet, but its power is uncertain.
The total radio luminosity at low frequency makes both of them
belong to the FR II radio galaxy class (see Fig. \ref{fig:p_core_tot}).
\jsix\ has been detected previously in VLBI observations by \cite{wang17}, but the
resolution was not sufficient to clearly see a collimated inner jet, leading the authors to
assign this emission to some outflow from the accretion disc or to 
plasma heated by the AGN X-ray corona.
However, our VLA observations clearly show the presence of a jet, implying that 
powerful jets (of the FR II type) can be present even in radio-quiet objects.

For both sources our modelling corresponds  to
two extremes: First, we assumed that the jets of all jetted sources are alike,
that they have similar bulk Lorentz factors, and
that their power is of the same order of their accretion disc luminosities.
Second, we instead assumed that their jet is only mildly relativistic.
We note that any solution in between these two extremes might be possible, as long
as the extended component energized by the power carried by the jet is 
compatible with the data we observe.
We have seen that the power required by the lobe of \jfive\ is of the same order
of the entire power of the `weak' jet, suggesting that this extreme solution is
not tenable.
But we cannot exclude a priori that its jet is somewhat less powerful than the
ones we see in blazars. 

The compactness of the radio structures could suggest that the sources belong to the GPS/CSS population \citep{odea21}, particularly in the case of \jfive. 
However, the expected young age of the GPS/CSS population contrasts with the very large black hole masses hosted in our sources (indicating a relatively `old' age), which requires the radio phase to not occur simultaneously with strong accretion phases. 
We also considered the hypothesis of a strong wind responsible for \jfive\ properties, but its radio luminosity is on the higher extreme of the range of radio powers associated with strong wind emitters. 
We also considered that such strong winds are always accompanied by significant nuclear obscuration \citep[as also described by][]{hwang18}, a feature that \jfive\ is completely lacking.

Independently of this, our modelling shows that even the `weak jet' solution
implies a jet power larger than a few $\times 10^{45}$ erg s$^{-1}$. 
This is larger than the power of many blazars studied in \cite{ghisellini14b}.

We can thus conclude that the jets in these sources are in any case crucial components of their structure, dynamics, and physics despite the fact that these sources can be formally classified as radio-quiet.

\section{Jets in the early Universe}
\label{sec:disc}

In the previous sections we derived the presence of jets in all three sources, 
and in particular we explored different options of jet orientation for \jsix{} and \jfive. 
We have concluded that despite the uncertainty on jet features, these two extremely massive sources
host relativistic jets even if they are classified as `radio-quiet' according to the 
standard definition. 
We selected them for this work only on the basis of their extreme masses: 
\jsix\ and \jfive\ are the two most massive quasars at $z>5$ and the only two with $M>10^{10}M_\odot$ 
at such high redshift.
What does finding jets in the most massive quasars at $z>5$ imply for the formation and evolution of the accreting black hole population?
\vskip 0.2cm

\begin{figure}
   \centering
   \includegraphics[width=\hsize]{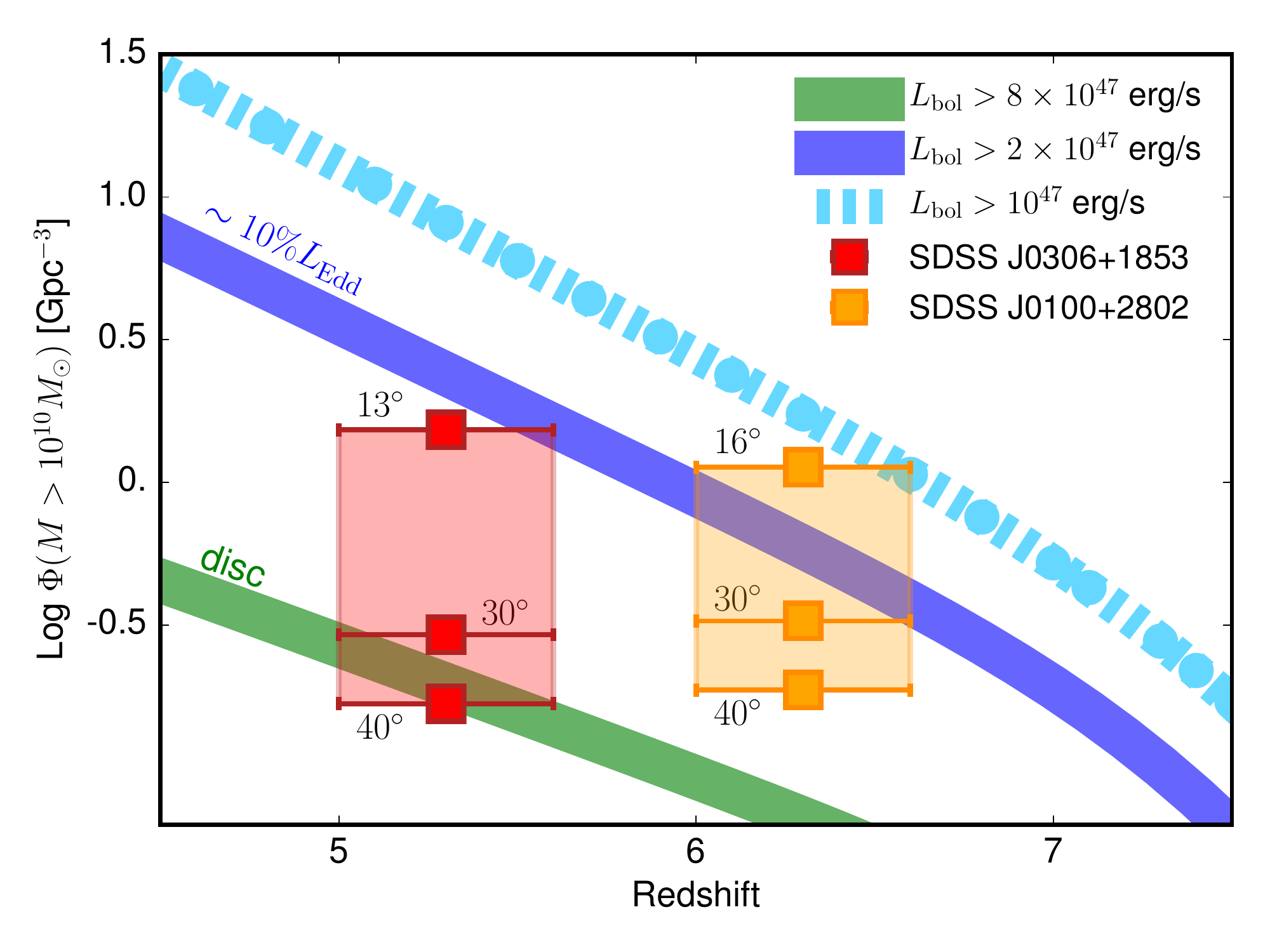}
   \caption{Co-moving number density of the jetted population associated with \jsix{} and \jfive{} (orange and red squares, respectively), 
            compared to the space density of quasars with $L_{\rm bol}>10^{47}$erg/s, $2\times10^{47}$erg/s, and $8\times10^{47}$erg/s 
            from \citet[][dashed light blue, solid blue, and solid green lines]{shen20}. 
            The densities extracted from the two sources are calculated for viewing angles of $40^\circ$ and $30^\circ$, and the smallest angles,            i.e.\ $13^\circ$ for \jfive\ and $16^\circ$ for \jsix, are derived from SED fitting.
            According to these estimates, the whole quasar population is hosted in jetted AGNs at $z>5$. 
            }
    \label{fig:density}%
    \end{figure}

Assuming that jetted quasars are distributed with uniform orientations in the sky and 
assuming one knows the viewing angle, one can estimate how many quasars 
with the same intrinsic features (mass, accretion rate, jet, etc.) as \jfive\ and \jsix\ exist at the 
same redshift. 
Specifically, one can infer the population size, $R_{10^{10}M_\odot}$, of differently aligned analogues of a source by knowing its viewing angle, assuming that it covers a solid angle, $\Delta\Omega$, and that the whole population is uniformly oriented in the sky. Eventually, this must be corrected for the selection sky area of the source. 
Each detected source is thus representative of a population of other $R_{10^{10}M_\odot}$ defined as:
\begin{equation}
    R_{10^{10}M_\odot} = \frac{A_{\rm sky}}{A_{\rm SDSS+FIRST}} \frac{4\pi}{\Delta\Omega}.
\end{equation}
We used this approach to evaluate how \jfive{} and \jsix{} can trace the population of $10^{10}M_\odot$ active black holes at $z=5.3$ and 6.3, respectively. 
Figure \ref{fig:density} shows the results for the discussed viewing angles $\theta=13^\circ$ and $16^\circ$, along with the aperture angle expected for an obscuring toroidal dust structure, $\theta\sim40^\circ$ (i.e.\ the maximum viewing angle expected for a Type 1 quasar), and an average viewing angle for this type of sources, $\theta\sim30^\circ$.
Specifically, this figure shows the co-moving number densities of $M>10^{10}M_\odot$ active black holes hosted in jetted sources at $z=5.3$ and $6.3$, as traced by our two quasars. 
We compared them with the corresponding curves for the whole AGN population with masses larger than $10^{10}M_\odot$ obtained from the luminosity function by \citet[]{shen20}. 
These authors base their study on an observational compilation that includes quasar observations in the rest-frame IR, B band, UV, and soft and hard X-ray performed in the past few decades, without explicit separation between jetted and non-jetted sources. 
We thus assumed that their bolometric luminosity function describes the whole AGN population, including jetted sources. 
We chose to compare our results with two different luminosity ranges from the \citet[]{shen20} results: a bolometric luminosity comparable to or higher than that of \jsix{} and \jfive{}, and a bolometric luminosity corresponding to (or higher than) $\sim10\%$ of the Eddington luminosity in the case of $10^{10}M_\odot$. 
The bolometric luminosities considered by \citet[]{shen20} were obtained through a phenomenological so-called bolometric correction commonly used in the literature to estimate the overall thermal output of the nuclear region of an AGN. Not only does it include the accretion emission from the disc, but it also adds the re-emitted IR  luminosity from the dusty obscuring structure and X-ray emitting corona. On average, the bolometric luminosity should be $L_{\rm bol}\simeq 2f(\theta,a)L_{\rm d}$, where $f(\theta,a)$ is the viewing angle pattern emission. 

Figure \ref{fig:density} clearly highlights a problematic feature: The population traced by \jfive{} and more dramatically by \jsix\ might outnumber what is expected from studying known catalogues thought to be relatively complete. Their co-moving number density is larger than the space density of not only their non-jetted counterparts at same luminosity range, but also of the overall known population, irrespective of the chosen viewing angle. 
The contrast is less extreme if the $10^{10}M_\odot$ population is assumed to be traced by $>10\%L_{\rm Edd}$ accreting sources (i.e.\ slower than \jsix{} and \jfive). 
Even in this case, though, the jetted sources appear to be much more numerous in comparison to non-jetted ones than the expected 10\% of the population observed at lower redshift, and the jetted ratio appears to increase going towards higher redshift. 
Jets clearly have a much more prominent presence and perhaps a more prominent role in the early Universe.

   \begin{figure}
   \hskip -0.4 cm
   \includegraphics[width=1.15\hsize]{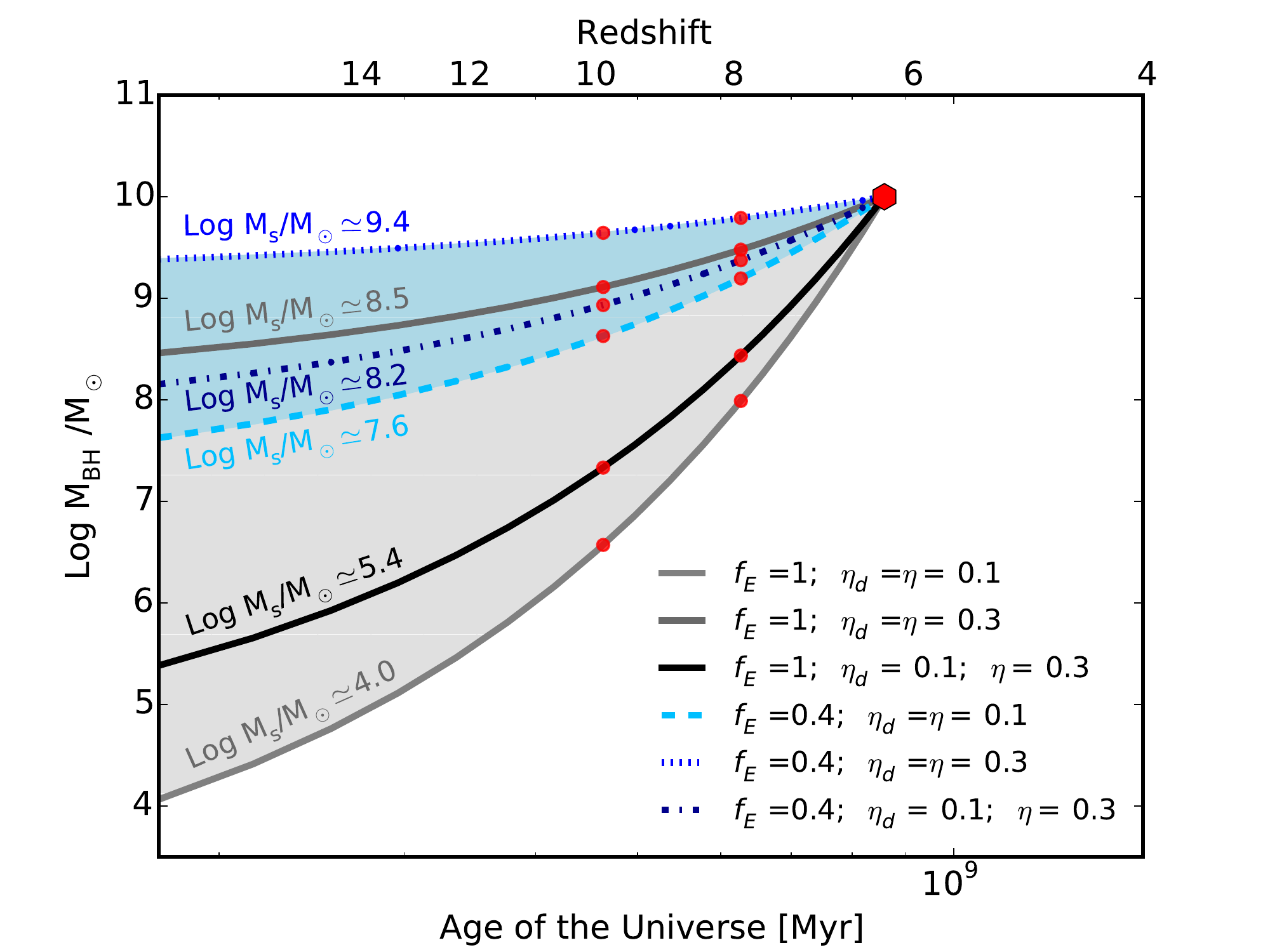}
   \caption{Black hole mass evolution of a black hole with $M=10^{10}M_\odot$ at $z=6.3$,
        assuming a continuous accretion on a Salpeter timescale at the Eddington limit 
        (solid lines) and at $f_{\rm E}=L_{\rm d}/L_{\rm Edd}=0.4$, i.e.\ close to the Eddington ratios derived for the two sources following the KERRBB approximation for a maximally spinning black hole (dotted, dashed, and dot-dashed lines). 
        The colours refer to different accretion efficiencies and combinations, as indicated in the legend. 
        Next to each line, the labels show the mass values of the hypothetical seeds located at $z=20$ for each accretion setup.
                }
    \label{fig:salp}%
    \end{figure}

Jetted sources that host extremely massive black holes are not easy to reconcile with their extremely high redshift. 
In fact, the presence of a jet is generally associated with a highly spinning accreting black hole, which corresponds to a higher radiative efficiency in the accretion process. 
In other words, a larger fraction of the accreting mass energy is radiated away compared to the case of a non-spinning black hole. 
Thus, less matter would ultimately accrete onto the central black hole, slowing down the accretion process itself. 
The standard radiative efficiency of $\eta=0.1$ is generally associated with a Newtonian approximation of a non-spinning black hole, while a maximally spinning object implies a radiative efficiency of $\eta=0.3$ \citep{thorne74}.
A slow accretion implies an extremely massive seed black hole, which can become problematic in the view of seed formation theories \citep{volonteri12}.

Assuming a Salpeter timescale \citep{salpeter64} for a continuous accretion from $z\simeq20$, we are able to estimate the seed black hole mass from which accretion started, as we show for different accretion and radiative efficiency combinations in Fig. \ref{fig:salp}. 
A $10^{10}M_\odot$ black hole accreting with an Eddington ratio of $f_{\rm E}=L_{\rm d}/L_{\rm Edd}\simeq40\%$ (as we derived in Sect. \ref{sec:accr}) up to $z=6.3$ needs a seed black hole with ${\rm Log}M_{\rm S}/M_\odot\simeq9.4$ if it is maximally spinning.

Even an accretion at the Eddington limit would not allow the presence of a Pop III star origin  to be inferred for the seed black hole ($M\sim10^2-10^3M_\odot$): With the same assumptions, the very presence of \jsix\ at $z=6.3$ implies a seed of ${\rm Log}M_{\rm S}/M_\odot\simeq8.5$.
A solution that would lead to a smaller seed black hole is that not all the gravitational energy released during the accretion process gets radiated away, only a fraction of it. 
As an example, we can assume, as in \cite{ghisellini13}, that a jetted SMBH releases a fraction of the accreting mass energy $\eta=0.3$, but only $\eta_{\rm d}=0.1$ gets radiated away. For a fixed observed luminosity, more mass is directly accreted onto the central black hole. The resulting $\eta_{\rm j}=\eta-\eta_{\rm d}$ would instead be involved in the jet launching. 
Under this assumption, for $f_{\rm E}=0.4$ the seed black hole needed would be less massive, albeit not drastically so. 
Nevertheless, the presence of a jet accelerates the accretion process, easing the problematic formation of extremely massive black holes such as the ones we are studying. 

More significantly, there was no initial clue that these sources host a jet: no radio emission and/or a  very low radio-loudness. 
The main issue is thus that, at very high redshift, extremely massive sources might not result as jetted according to large surveys, but they might be hiding powerful relativistic jets in plain sight. 
High-resolution radio observations are definitely needed in order to observe their jets and derive their features. To explore whether this covers all massive sources at $z>5$, a more extensive search with VLA- or VLBI-dedicated observations is necessary in order to overcome the limitations of microJansky-limited radio surveys.

\section{Conclusions}
\label{sec:concl}

We performed VLA observations of three $z>5$ quasars, \jsix, \jfive,\ and \b2, 
in order to look for jet signatures in their radio emission. 
The first two sources are radio-quiet quasars, and as such they are expected to not have a jet, 
while the last is strongly radio-loud and a well-known jetted source.
We then studied their broadband emissions in order to derive accretion and jet features 
and ultimately investigate the presence and details of relativistic jets 
in the early Universe that coexisted with extremely massive quasars. 

B2~1023+25 was already classified as a high-redshift blazar thanks 
to its X-ray features and existent VLBI observations. 
Our data are consistent with the previous classifications, showing a compact, 
bright radio emission that is consistent with a powerful relativistic jet aligned to our line of sight. 

Previous observations of \jsix\  in the radio band hint towards an AGN emission for its 
radio flux, but we managed to resolve its VLA emission, which shows a core and a 
slightly extended emission. 
\jfive\ was previously undetected in the radio band. We observe strong, unresolved radio emission 
at both 1.4 and 5GHz. 
Even though their radio-loudness does not describe them as typical jetted sources, 
their radio power is clearly consistent with that of FRII sources. 
We thus can conclude that they host relativistic jets, even if their extreme optical brightness 
makes them radio-quiet. 
This reinforces the idea suggested by \cite{padovani16} that the radio-quiet versus\ radio-loud dichotomy should not uniquely correspond to non-jetted versus\ jetted classifications.

The broadband SED fitting of \jfive\ and \jsix\ is consistent with the presence of a powerful relativistic jet with viewing angles of $13^\circ$ and $16^\circ$, respectively, and a standard bulk Lorentz factor of $\Gamma=13$ in both cases. 
Both SEDs are also consistent with larger viewing angles ($\sim30^\circ$), which are associated with a smaller Lorentz factor of 2. 
We nevertheless prefer the solution with a smaller viewing angle since the injected power in the hypothetic lobe structures should be a substantial fraction (if not almost all) of the total jet power in the case of a slower solution, leaving no room for dispersion.

We note that \jsix\ and \jfive\ were only selected on the basis of their mass: They were the only two quasars at $z>5$ with masses larger than $10^{10}M_\odot$. 
We did not have radio-based suggestions that these sources could host jets. 
Their extreme masses, associated with strong jets, challenge the formation and evolution picture of the first SMBHs: There is not enough time to build them according to standard models. 
The presence of jets in both of them suggests that they might be playing an interesting role in the fast formation of extremely massive sources: They might even be facilitating extreme events of accretion in the early Universe and be much more present in the massive black hole population at high $z$. 

While jets are commonly searched for in radio surveys, we have proven in this work that this standard approach might lead to a substantial underestimate of their presence at high $z$. 
Slightly misaligned jets in AGNs can only be detected thanks to deep radio observations: microJansky-level surveys such as FIRST are not efficient in detecting them. 
Moreover, another issue related to their detection is linked to the apparent absence of a strongly emitting radio lobe. 
Extended radio lobes are known to be quenched by the interaction of electrons with the CMB radiation, which is much more energetically dense at $z>3$. 
Even at lower redshift, it can be difficult to observe the extended emission in some cases. 
The powerful and widely studied blazar 3C~273 had lacked extended lobe radio observation until a few years ago due to the dynamical range of radio interferometric observations \citep{perley17}. 
The eventual observations revealed a somewhat small lobe power, of $\sim3\times10^{41}$erg/s.

The detection of extended radio emission and relativistically beamed emission from the core jet of very high-redshift sources might be more difficult than expected, but jets might be much more frequent in the extremely massive quasar population at $z>5$, suggesting a more prominent role in the early formation of these sources. 
An extensive search for jets therefore requires high-resolution radio observations.

\begin{acknowledgements}
We thank the referee for their valuable comments and suggestions. The National Radio Astronomy Observatory is a facility of the National Science Foundation operated under cooperative agreement by Associated Universities, Inc.
\end{acknowledgements}


\bibliographystyle{aa}
\bibliography{BIB} 

\end{document}